\documentclass[twocolumn,preprintnumbers,amsmath,amssymb,10pt]{revtex4}
\usepackage{amsmath}
\usepackage[ruled]{algorithm2e}
\usepackage{graphicx}
\usepackage{bm}
\usepackage[all]{xy}
\usepackage{color}
\usepackage{bbm}
\usepackage{amssymb}
\usepackage{extarrows}
\begin{document}
\title{Negative Shannon Information Hides Networks}

\author{Ming-Xing Luo}

\affiliation{\small{} School of Information Science and Technology, Southwest Jiaotong University, Chengdu 610031, China;
\\
CSNMT, International Cooperation Research Center of China, Chengdu 610031, China;
}

\newtheorem{Thm}{Theorem}
\newtheorem{Def}{Definition}
\newtheorem{Prop}{Proposition}
\newtheorem{Lem}{Lemma}
\newtheorem{Cor}{Corollary}
\newtheorem{Exa}{Example}
\newtheorem{Alg}{Algorithm}

\begin{abstract}
Shannon information was defined for characterizing the uncertainty information of classical probabilistic distributions. As an uncertainty measure it is generally believed to be positive. This holds for any information quantity from two random variables because of the polymatroidal axioms. However, it is unknown why there is negative information for more than two random variables on finite dimensional spaces. We first show the negative tripartite Shannon mutual information implies specific Bayesian network representations of its joint distribution. { We then show that the negative Shannon information is obtained from general tripartite Bayesian networks with quantum realizations.} This provides a device-independent witness of negative Shannon information. We finally extend the result for general networks. The present result shows new insights in the network compatibility from non-Shannon information inequalities.
\end{abstract}
\maketitle

\section{Introduction}

For a given discrete Markov process a fundamental problem is how to characterize the information produced in such a statistical process. Suppose the statistical outcomes of a set of possible events $\{x_1,\cdots, x_n\}$ with respectively occurrence frequencies as $p_1, \cdots, p_n \in [0,1]$. Is there a measure of how uncertain is of the outcome except for its distribution? For any such a quantity denotes as $H(p_1, p_2, \cdots, p_n)$, it is reasonable to satisfy the following axioms: (1) $H$ is continuous function in each variable $p_i$ of the probability. (2) $H$ is an increasing function of sample number for the uniform distribution. (3) $H$ is weighted summation of its single values if any one choice is changed into two. These axioms imply the unique entropy given by Shannon \cite{Shannon} as
\begin{eqnarray}
H(p_1, p_2, \cdots, p_n)=-\sum_{i=1}^np_i\log{}p_i
\label{eqn1}
\end{eqnarray}
This features the average uncertainty of a given statistical process.

\begin{figure}
\begin{center}
\resizebox{220pt}{300pt}{\includegraphics{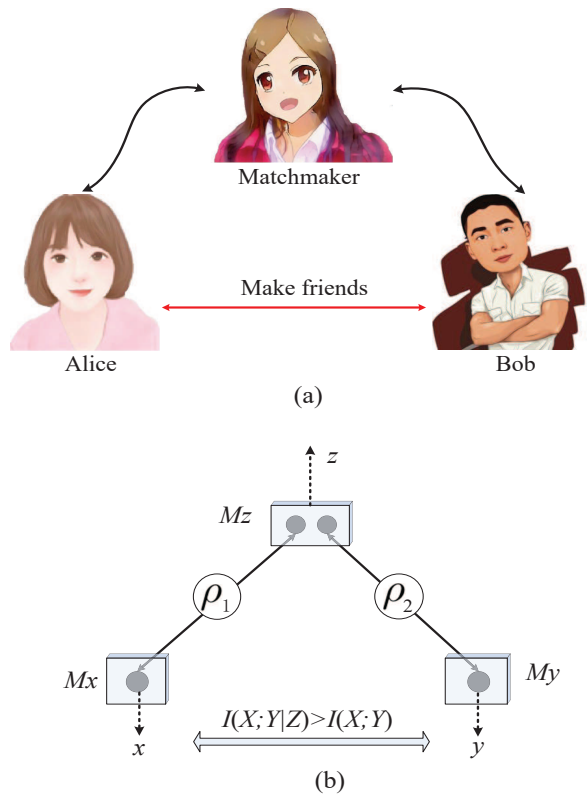}}
\end{center}
\caption{\small (Color online) Schematic network configuration compatible with negative Shannon mutual information. (a) A classical semantic example. Alice and Bob who have not shared any relationship make friends assisted by Matchmaker who is familiar with both. (b) A quantum realization of Bayesian network in (a). There are two entangled states \cite{EPR} $\rho_1$ and $\rho_2$ which are shared by three parties,  where $\rho_1=\rho_2=\frac{1}{\sqrt{2}}(|00\rangle+|11\rangle)$ on two-dimensional Hilbert space $\mathcal{H}$ spanned by the orthogonal basis $\{|0\rangle, |1\rangle\}$. Under local quantum measurements with positive-operator-value matrices $\{M_x=|x\rangle\langle x|,x\in \{0,1\}\}, \{M_y=|y\rangle\langle y|,y\in \{0,1\}\}$ and $\{M_z=|z\rangle\langle z|,z\in \{0, 1\}\}$, three parties can generate a joint distribution $P_{xyz}=\frac{1}{4}[000]+\frac{1}{4}[011]+\frac{1}{4}[101]+\frac{1}{4}[110]$. This follows $I(X;Y)=0$ and $I(X;Y|Z)=1$. This means Alice and Bob who are initially independent of each other (e.g., $I(X;Y)=0$) can build new correlations conditional on local measurements of Matchmaker (e.g., $I(X;Y|Z)=0$).}
\label{fig1}
\end{figure}

The Shannon entropy shows a remarkable application for mutual information of two discrete random variables as
\begin{eqnarray}
I(X;Y)=H(X)+H(Y)-H(X,Y)
\label{eqn1a}
\end{eqnarray}
where the mutual information means the uncertainty between two variables $X$ and $Y$. It is a measure of the information to which knowledge of one variable reduces uncertainty about the other. These entropy functions satisfy the polymatroidal axioms \cite{Fuji} of $H(X), I(X;Y)\geq 0$ for any finite dimensional variables $X$ and $Y$. Shannon actually shows general information inequalities which are the "physical laws" for characterizing the fundamental limits in classical communications and compression \cite{Shannon}. However, this intrigues a surprising feature of negative information for three or more discrete variables beyond the polymatroidal axioms. Each one in fact presents a so-called non-Shannon inequality \cite{Zhang1997}. One primitive example is from the mutual information contained in three discrete variables $X, Y$ and $Z$ on finite dimensional spaces as
\begin{eqnarray}
I(X;Y;Z)&=&I(X;Y)-I(X;Y|Z)
\nonumber
\\
&=&H(X)+H(Y)+H(Z)+H(X,Y,Z)
\nonumber
\\
&&-H(X,Y)-H(Y,Z)-H(X,Z)
\label{eqn2}
\end{eqnarray}
where $I(X;Y|Z)=H(Y,Z)+H(X,Z)-H(Z)-H(X,Y,Z)$ denotes the mutual information conditional on the outcome of variable $Z$. One example is shown in Fig.\ref{fig1}.  The proper local measurements on the network (a) may generate a joint probability distribution $P_{xyz}$ with the notation of  $P_{xyz}=\frac{1}{4}[000]+\frac{1}{4}[011]+\frac{1}{4}[101]+\frac{1}{4}[110]$, where $[xyz]$ denotes joint event $(X=x,Y=y,Z=z)$, and the probability of joint outcome $[xyz]=[000], [011], [101]$ or $[110]$ is $\frac{1}{4}$. This probability distribution yields to a negative information of $I(X;Y;Z)=-1$. Especially, the new correlations are built for two independent parties Alice and Bob assisted by the other's local operations, that is, the mutual information conditional on the outcome of $Z$ is given by $I(X;Y|Z)=1$ while the mutual information $I(X;Y)$ is zero. This may imply a simple explanation of the Shannon negative information from Bayesian networks \cite{Pearl}. This intrigues a natural problem for characterizing general negative Shannon information.

In Hilbert space formulation of Bayesian network in Fig.\ref{fig1}(a), each bipartite edge is replaced by an entanglement \cite{EPR,NC}, as shown in Fig.\ref{fig1}(b). Under the proper postulates of quantum state representation, quantum measurement, Born rule and tensor decomposition of compose systems, the quantum probability is given by $p(x,y,z)={\rm Tr}((M_x\otimes M_y\otimes M_z)\rho_1\otimes \rho_2)$, where $\{M_x\}$, $\{M_y\}$, and $\{M_z\}$ denotes respective quantum measurements of Alice, Bob and Matchmaker, and ${\rm Tr}(\cdot)$ denotes the trace operation of matrix. This quantum probability shows not only the similar features of classical statistics of $I(X;Y;Z)\leq 0$, but also the quantumness of entanglement \cite{EPR,Bell}, i.e., two independent parties Alice and Bob can build quantum entanglement assisted by the other's local operations and classical communication \cite{Entswap1993,Cava}. This provides a simple physical model for verifying Bayesian network and  the new quantumness of entanglement assisted by other party from its statistical distribution.

Our motivation in this work is to investigate a general problem of the Bayesian network compatibility of the negative Shannon mutual information. For a given tripartite joint probability distribution with negative mutual information on finite sample spaces, we firstly classify all the compatible Bayesian networks. We show there are intrinsic network configurations for these distributions, that is, chain network consisting of two edges or triangle network consisting of three edges. This implies a device-independent verification of negative Shannon mutual information using quantum networks. The feature of negative Shannon mutual information is generic for any tripartite quantum entangled network or general multipartite networks.

\section{Result}

\subsection{Negative Shannon mutual information in Bayesian network model}.

\begin{figure}
\begin{center}
\resizebox{170pt}{180pt}{\includegraphics{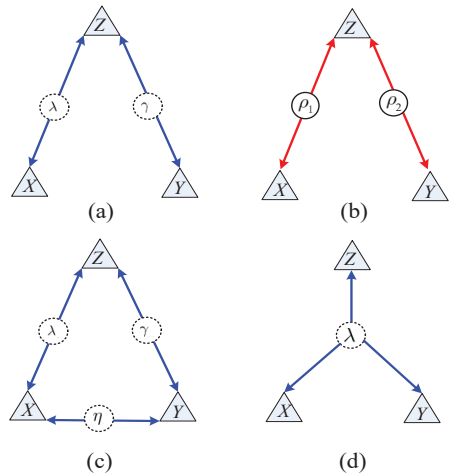}}
\end{center}
\caption{\small (Color online) Schematic Markov compatibility of negative Shannon mutual information. (a) Classical chain network consisting of two latent variables $\lambda$ and $\gamma$. (b) Quantum chain network consisting of two bipartite states $\rho_1$ and $\rho_2$ on Hilbert space $\mathcal{H}$. (c) Classical triangle network consisting of three latent variables $\lambda$, $\gamma$ and $\eta$. (d) Classical tripartite network consisting of one latent variable $\lambda$.}
\label{fig2}
\end{figure}

We first introduce some notations of { Bayesian } networks \cite{Pearl}. A graph $\mathcal{G}$ consists of a vertex (or node) set $V$, and an edge (or link) set $E$. The vertices in a given graph are corresponding to measurable variables, and the edges denote certain relationships that hold in pairs of variables. A bi-directed edge denotes the existence of unobserved common causes. These edges will be marked as curved arcs with two arrowheads, as shown in Fig.\ref{fig2}. If all edges are directed, we then have a directed graph.

Directed graph may include directed cycles. One example is given by $X \to Y\to Z\to X$ with $X, Y, Z \in V$, which represents mutual causation or feedback processes. The self-loops (e.g., $X \to X$) are not allowed in what follows. A graph that contains no directed cycle is called acyclic. A graph that is both directed and acyclic is called a directed acyclic graph (DAG). A family in a graph is a set of nodes containing a node and all its parents, where the parents of one node mean all nodes which are connected to it.

Denote a probability distribution as $P_{x}=\tilde{\Sigma}_{x}p_{x}[x]$ with random variable $X$ on a finite sample space $\mathcal{X}$, where $[x]$ denotes the event of $X=x$ and $p_{x}$ denotes the probability of the outcome $x$. The notation $\tilde{\Sigma}$ does not mean the summation but a notation of union of all possible events of a given probability distribution. Similar notations will be used for multivariate joint distributions $P_{x_1\cdots{}x_n}$ on finite sample space $\times_{i=1}^n\mathcal{X}_i$. Consider the task of specifying an arbitrary joint distribution $P_{x_1\cdots{}x_n}=\tilde{\Sigma}_{x_1\cdots{}x_n}p_{x_1\cdots{}x_n}[x_1\cdots{}x_n]$ for $n$ random variables, $X_1, \cdots, X_n$ on finite sample spaces $\mathcal{X}_1, \cdots, \mathcal{X}_n$, respectively, and $[x_1\cdots{}x_n]$ denotes the joint event of $X_1=x_1, \cdots, X_n=x_n$. The basic Bayes rule allows us decompose $p_{x_1\cdots{}x_n}$ into
\begin{eqnarray}
p_{x_1\cdots{}x_n}=\prod_jp_{x_j|x_1,\cdots, x_{j-1}}
\label{eqn8a}
\end{eqnarray}
where $p_{x_j|x_1,\cdots, x_{j-1}}$ denotes the probability of outcome $x_j$ conditional on the outcomes of predecessors $x_1,\cdots, x_{j-1}$. Suppose that each $x_j$ is dependent of a small subset $pa(x_j)$ of its predecessors. We have the following definition.

\textbf{Definition 1}. (Markovian Parents) \cite{Pearl} Let $V= \{X_1,\cdots, X_n\}$ be an ordered set of measurable variables, and let $P_{x_1, \cdots, x_n}$ be the joint probability distribution on these variables. A set of variables $PA_j$ is said to be Markovian parents of $X_j$ if $PA_j$ is a minimal set of predecessors of $X_j$ that renders $X_j$ independent of all its other predecessors, that is,
\begin{eqnarray}
p_{x_j|pa_j}=p_{x_j|x_1,\cdots,x_{j-1}}
\label{A1}
\end{eqnarray}
and such that no proper subset of $PA_j$ satisfies (\ref{A1}).

Definition 1 implies for each $X_j$ there is a set $PA_j$ of preceding variables for determining its probability. This can be represented by DAG, where $PA_j$ denotes all the parent nodes toward the node $X_j$ \cite{Pearl}. Definition 1 provides a simple recursive algorithm for constructing such a DAG for a given $P_{x_1\cdots{}x_n}=\tilde{\sum}_{x_1,\cdots, x_n}p_{x_1\cdots{}x_n}[x_1\cdots{}x_n]$ as follows.
\begin{itemize}
\item[]\textbf{Algorithm 1}
\item[(i)] Starting with the pair $(X_1, X_2)$, we draw an arrow from $X_1$ to $X_2$ if and only if the two variables are dependent.
\item[(ii)] For $X_3$, we draw an arrow from either $X_1$ or $X_2$ to $X_3$ if $X_3$ is dependent of $X_1$ or $X_2$;
\item[(iii)] For $j\geq 3$, one can select any minimal set $PA_j$ of $X_j$'s possible predecessors. And then, draw an arrow from each member in $PA_j$ to $X_j$.
\end{itemize}

This follows an iterative algorithm to get a DAG of Bayesian network compatible with the given distribution $P_{x_1\cdots{}x_n}$. It { has been shown } that $PA_j$ is unique for a given distribution $P_{x_1\cdots{}x_n}$ \cite{Pearl}. From the Reichenbach's common cause principle \cite{Pearl,Markov}, it allows a Markovian decomposition as
\begin{eqnarray}
p_{x_1\cdots{}x_n}=\prod_jp_{x_j|pa(x_j)}
\label{eqn8b}
\end{eqnarray}
The Markovian dependence is represented by the directed acyclic graph (DAG) of Bayesian networks \cite{Pearl}.

The DAG shows probabilistic and statistical importance for data mining and efficient inferences. A basic problem in statistics theory is to explore the related DAG for a given statistical distribution. The most common way to explore possible DAGs from observations is based on the Markov decomposition { in Eq.(\ref{eqn8b})} and the faithfulness assumption \cite{Pearl,SG}.

\textbf{Definition 2} \cite{Pearl} (Markov Compatibility) If a probability function $P_{x_1\cdots{}x_n}$ admits the factorization of (\ref{eqn8b}) relative to DAG $\mathcal{G}$, then $\mathcal{G}$ and $P_{x_1\cdots{}x_n}$ are compatible.

In classical realization of Bayesian networks, each edge is represented by one measurable variable on proper measurable space, and each outcome depends on all the related variables \cite{Pearl}. The joint distribution is a multivariate function of all outcomes. This allows us to decompose $P_{x_1\cdots{}x_n}$ with one measurable variable $\lambda$ \cite{Bell} as:
\begin{eqnarray}
p_{x_1\cdots{}x_n}=\int_{\Omega}p(x_1|\lambda)\cdots {}p(x_n|\lambda)
\mu(\lambda)d\lambda
\label{eqn8}
\end{eqnarray}
where $(\Omega,\mu(\lambda))$ denotes the measurable space of unobservable latent variable $\lambda$, and $\mu(\lambda)$ denotes the probability of $\lambda$, and $p(x_n|\lambda)$ denotes the characteristic function of outcome $x_n$ conditional on the variable $\lambda$. This kind of Bayesian networks with latent variables shows non-trivial constraints on its correlations \cite{TP,KT}.

For a joint distribution of two random variables $X$ and $Y$ on finite sample spaces $\mathcal{X}\times \mathcal{Y}$, { Eq.(\ref{eqn1})} is used for featuring the common uncertainty of both variables \cite{Shannon}. The nonnegative of $H$ and $I$ from the polymatroidal axioms \cite{Fuji} is useful for solving the Markov compatibility with single latent variable { in Eq.(\ref{eqn8})}. Instead, the entropy function { in Eq.(\ref{eqn2})} shows the tripartite mutual uncertainty only for Markov chains \cite{Markov}, that is,
\begin{eqnarray}
I(X;Y;Z)\geq0
\label{Markov}
\end{eqnarray}
if $\{X,Y, Z\}$ (under any order) consists of a Markov chain \cite{Pearl} (see proofs in Appendix A). This inspires a generalized unordered Markov condition of $I(X;Y;Z)\geq 0$. The joint probability distribution can be generated from one single latent variable as Eq.(\ref{eqn8}). Remarkably, there are joint distributions implying negative Shannon mutual information, that is, $I(X;Y;Z)<0$. One example is shown in Fig.\ref{fig2}(a). The coarse-grained single-variable model (\ref{eqn8}) does not imply any intrinsic feature of this case. Instead, we prove new Markov compatibilities for these distributions using Bayesian networks \cite{Pearl}.

\textit{Case one}. From the definition { in Eq.(\ref{eqn2})} the first scenarios satisfies the following constrains:
\begin{eqnarray}
I(X;Y)&=&0, I(X;Y|Z)>0.
\label{eqn9}
\end{eqnarray}
Different from the Markov conditional independence of $p(x,y|z)=p(x|z)p(y|z)$ \cite{Pearl}, the present condition { in Eq.(\ref{eqn9})} implies a conditional dependence, that is, two independent random variables $X$ and $Y$ on $\mathcal{X}\times \mathcal{Y}$ can build new correlations conditional on the outcome of variable $Z$. It can be mathematically formulated as $p(x,y)=p(x)p(y)$ and $p(x,y|z)\not=p(x|z)p(y|z)$. The so-called anti-Markov condition provides a primitive explanation of nonnegative Shannon mutual information with $I(X;Y;Z)<0$.

Especially, consider the example shown in Fig.\ref{fig2}(a). Combined with classical Birkhoff transformation \cite{Birk} (e.g., a doubly stochastic matrix $(a_{ij})$ which satisfies each column or each row consists of a probability distribution, i.e., $\sum_i a_{ij}= \sum_ja_{ij}=1$ and $ a_{ij}\geq 0$), the joint distribution $P_{xyz}=\frac{1}{4}[000]+\frac{1}{4}[011]+\frac{1}{4}[101]+\frac{1}{4}[110]$ can be obtained from the following joint distribution of four variables as
\begin{eqnarray}
P_{xyz_1z_2}=\frac{1}{4}[0000]+\frac{1}{4}[0101]+\frac{1}{4}[1010]+\frac{1}{4}[1111]
\label{eqn11}
\end{eqnarray}
where the outcomes of the variable $Z$ are classically encoded as: $\mathcal{F}: z_1z_2\mapsto z=z_1\oplus {}z_2$. Different from the joint distribution $P_{xyz}$, the new distribution in Eq.(\ref{eqn11}) allows the following Markovian decomposition of  $P_{xyz_1z_2}=P_{xz_1}P_{yz_2}$, where $P_{xz_1}$ and $P_{yz_2}$ are joint distributions of two variables given by $P_{xz_1}=P_{yz_2}=\frac{1}{2}[00]+\frac{1}{2}[11]$. This means both the variables $X$ and $Y$ are independent. Moreover, it is easy to get $H(X)=H(Y)=1, H(X,Y)=H(X,Y,Z)=2$ and $H(X,Z)=H(Y,Z)=1$. This implies the distribution $P_{xyz_1z_2}$ satisfies the inequality (\ref{eqn9}).

Now, we continue the proof. Combining with { Eq.(\ref{eqn8})} the distribution $P_{xyz}$ allows a classical chain network decomposition as shown in Fig.\ref{fig2}(a). This can be represented by
\begin{eqnarray}
p_{xyz}&=&\int_{\Omega_1\otimes \Omega_2} p(x|\lambda)p(y|\gamma)p(z|\lambda,\gamma)
\nonumber
\\
&&\times \mu(\lambda)\mu(\gamma)d\lambda{}d\gamma
\label{eqn12a}
\end{eqnarray}
where $(\Omega,\mu(\lambda))$ denotes the measurable space of the latent variable $\lambda$, and $\mu(\lambda)$ denotes the probability of $\lambda$, and similar definitions for the latent variable $\gamma$. This example can be extended for general probability distribution satisfying the condition in Eq.(\ref{eqn9}). Interestingly, a further analysis shows the rigidity of this Bayesian network compatibility with any joint distribution satisfying the condition { in Eq.(\ref{eqn9})} (see Appendix B).

\textbf{Result 1}. Any joint distribution satisfying the condition in Eq.(\ref{eqn9}) is compatible with a chain Bayesian network.

In Hilbert space formulation, a finite-dimensional pure state is represented by a normalized vector $|\phi\rangle$ in Hilbert space $\mathcal{H}_A$ \cite{NC}. An ensemble of pure states $|\phi_i\rangle$ with a mixing probability $p_i$ is represented by a density matrix $\rho=\sum_ip_i|\phi_i\rangle\langle \phi_i|$ on Hilbert space $\mathcal{H}_A$. Here, $\rho$ is positive semidefinite matrix with unit trace. The multipartite quantum system is defined on the tensor of local states, i.e., the tensor of Hilbert space as $\otimes_{i=1}^n\mathcal{H}_{A_i}$. Any measurement acting on $\mathcal{H}_A$ consists of an ensemble $\{M_{x_i}\}$ of projection operators or generalized positive semidefinite operators satisfying $\sum_{x_i}M_{x_i}=\openone_{A_{i}}$ with the identity operator $\openone_{A_i}$. After all the local measurements on a given state $\rho$ on Hilbert space $\otimes_{i=1}^n\mathcal{H}_{A_i}$, from Born rule, the quantum joint probability is given by
\begin{eqnarray}
p_{x_1\cdots x_n}={\rm Tr}[(M_{x_1}\otimes \cdots \otimes M_{x_n}) \rho]
\end{eqnarray}
for the joint outcome $x_1, \cdots, x_n$.

For special case of two particles $A$ and $B$, a state $\rho$ on Hilbert space $\mathcal{H}_{A}\otimes \mathcal{H}_B$ is entangled if it cannot be decomposed into
\begin{eqnarray}
\rho=\sum_{i}p_i\rho_A^{(i)} \otimes \rho_B^{(i)}
\end{eqnarray}
where $\{p_i\}$ is a probability distribution, $\rho_A^{(i)}$ and $\rho_B^{(i)}$ are states of respective particle $A$ and $B$. Similar definitions may be extended for multiple particles \cite{HHH}.

Instead of classical chain network in Fig.\ref{fig2}(a), for any joint distribution $P_{xyz}$ satisfying the condition (\ref{eqn9}) it is compatible with quantum networks consisting of two generalized Einstein-Podolsky-Rosen (EPR) states \cite{EPR}: $|\phi_1\rangle_{AB}=\frac{1}{\sqrt{2}}(|00\rangle+|11\rangle)$ and $|\phi_2\rangle_{B'C}=\frac{1}{\sqrt{2}}(|00\rangle+|11\rangle)$, as shown in Fig.\ref{fig2}(b). A simple local measurement strategy implies a quantum joint distribution
\begin{eqnarray}
P_{xyz}=\frac{1}{4}[0000]+\frac{1}{4}[0101]+\frac{1}{4}[1010]+\frac{1}{4}[1111]
\label{eqn13b}
\end{eqnarray}
where Alice and Bob performs respective projection measurements $\{M_x, x\in \{0,1\}\}$ and $\{M_y, y\in \{0,1\}\}$ while the other performs local measurement $\{M_z=|z\rangle\langle z|, z\in \{00, \cdots, 11\}\}$. This provides a possible experimental verification of any negative Shannon information satisfying Eq.(\ref{eqn9}). As a directive result, any joint distribution { in Eq.(\ref{eqn13b})} generated from local measurements on the quantum chain network in Fig.\ref{fig2}(b) with any bipartite states $\rho_1$ and $\rho_2$ is compatible with the classical distribution { in Eq.(\ref{eqn12a})}, that is,
\begin{eqnarray}
\mathcal{S}_{c}=\mathcal{S}_{q}
\label{eqn13c}
\end{eqnarray}
where $\mathcal{S}_{c}$ consists of all classical probability distributions in Eq. (\ref{eqn12a}) with respect to any two measurable variables $\lambda$ and $\gamma$, or equivalently represented by $\mathcal{S}_{c}=\{P_{xyz}^{(c)}|I(X;Y)=0,I(X;Y|Z)>0\}$;  $\mathcal{S}_{q}$ consists of all the quantum probability distributions { in Eq.(\ref{eqn13b})} derived from any two quantum states $\rho_1$ and $\rho_2$, that is, $\mathcal{S}_{q}=\{P_{xyz}^{(q)}|I(X;Y)=0,I(X;Y|Z)>0\}$. The equality (\ref{eqn13c}) means that there is no quantum nonlocality beyond classical networks for quantum chain network if each party has only one set of local measurements. This means that for a tripartite chain network its classical realization with two independent variables can simulate all quantum correlations from any realization of two independent entangled states with one measurement setting per party. This is different from previous results with more than one measurement settings on chain networks \cite{Cava,15}. It is also different from the triangle network consisting of three variables with one measurement setting \cite{Marc2019}. This may inspire another interesting problem for what the network ingredients may inspire the quantum nonlocality for a general network.

\textit{Case two}. For dependent random variables $X$ and $Y$ on respective finite sample spaces $\mathcal{X}$ and $\mathcal{Y}$, it suggests the second scenarios for $I(X;Y;Z)\leq 0$ as
\begin{eqnarray}
&&I(X;Y)>0, I(X;Y|Z)>0.
\label{eqn13}
\end{eqnarray}
One example is shown as
\begin{eqnarray}
P_{xyz}&=&\frac{1}{8}[000]+\frac{1}{8}[011]+\frac{1}{8}[102]+\frac{1}{8}[113]
\nonumber
\\
&&+\frac{1}{8}[220]+\frac{1}{8}[231]+\frac{1}{8}[322]+\frac{1}{8}[333]
\label{eqn14}
\end{eqnarray}
which has $I(X;Y)=1$ and $I(X;Y|Z)=2$. Here, the conditional mutual information $I(X;Y|Z)$ is defined { in Eq.(\ref{eqn2})}. By using the binary representation of $i$, $g: i\mapsto i_1i_2$, we may construct a compatible network model \cite{Pearl} by two steps.  One is to split $P_{xyz}$ into two joint distributions of $P_{x_1y_1}=\frac{1}{2}[00]+\frac{1}{2}[11]$ and $P_{x_2y_2z}$ in Eq.(\ref{eqn11}) by using local classical transformation $g$, that is,
\begin{eqnarray}
P_{g(x)g(y)z}&=&P_{x_1x_2y_1y_2z}=P_{x_1y_1}\times P_{x_2y_2z}
\nonumber
\\
&=&\frac{1}{8}[00000]+\frac{1}{8}[00011]+\frac{1}{8}[01002]+\frac{1}{8}[01013]
\nonumber
\\
&&+\frac{1}{8}[10100]+\frac{1}{8}[10111]+\frac{1}{8}[11102]+\frac{1}{8}[11113]
\nonumber\\
&{ g^{-1} \atop \to }&P_{xyz}
\label{eqn15a}
\end{eqnarray}
under the inverse mapping $g^{-1}$ of $g$ for $x$ and $y$, where the joint distribution $P_{x_1y_1}$ satisfies $I(X_1;X_2)=I(X;Y)$, and the joint distribution $P_{x_2y_2z}$ satisfies $I(X_2;Y_2)=0$ and $I(X_2;Y_2|Z)=2$.  Note that the joint distribution $P_{x_1y_1}$ can be generated by one latent variable $\lambda_3$ from Eq.(\ref{eqn8}) while $P_{x_2y_2z}$ is compatible with the chain network using Result 1. This fact implies a new compatible network consisting of three latent variables, as shown in Fig.\ref{fig2}(c).

In general, for a given distribution $P_{xyz}$ satisfying the condition { in Eq.(\ref{eqn13})}, suppose there is classical transformation $g: x\mapsto x_1x_2$ such that
\begin{eqnarray}
P_{g(x)g(y)z}&=&P_{x_1x_2y_1y_2z}=P_{x_1y_1}\times P_{x_2y_2z},
\label{eqn15}
\end{eqnarray}
where the joint distribution $P_{x_2y_2z}$ satisfies the condition in Eq. (\ref{eqn9}). Under this assumption, we can obtain the following result.

\textbf{Result 2}. For any joint distribution on finite sample spaces satisfying the condition in Eq.(\ref{eqn13}), there is a compatible triangle network if the decomposition in Eq.(\ref{eqn15}) holds.

A recent result shows triangle quantum network shows nonlocal correlations beyond all classical realizations under local measurement assumptions \cite{Marc2019}, that is, by performing one set of local measurements on a triangle quantum network consisting of three entangled states there are quantum tripartite joint probability $P_{xyz}$ which cannot be generated from any classical triangle network consisting of three measurable variables and local measurements. Interestingly, all of their quantum distributions satisfy the condition { in Eq.(\ref{eqn13})}. This implies the { inequivalence} of quantum and classical realizations of triangle network in Fig.\ref{fig2}(c). Thus the decomposition { in Eq.(\ref{eqn15})} provides a sufficient condition to verify triangle network, and may be evaluated by using numeric methods \cite{Spirt}.

The other is a tripartite network consisting of one latent variable as shown in Fig.\ref{fig2}(d). One example is the distribution $P_{xyz}=\frac{1}{3}[001]+\frac{1}{3}[010]+\frac{1}{3}[100]$,  which cannot be generated from all networks in Fig.\ref{fig2}(a)-(c) \cite{Blou,NW,Dur}.

This yields to a further problem to distinguish different configurations of triangle networks. Here, we present an informational method as (see proofs in Appendix C):
\begin{eqnarray}
I(X;Y;Z)&\leq& \min\{H(X|Y,Z), H(Y|X,Z), H(Z|X,Y)\}
\nonumber
\\
\label{eqn16a}
\\
I(X;Y;Z)&>&\frac{3}{2}(H(X)+H(Y)+H(Z))
\nonumber
\\
&&-H(X,Y)-H(X,Z)-H(Y,Z)
\label{eqn16b}
\end{eqnarray}
which hold for the network in Fig.\ref{fig2}(c). Instead, for the network in Fig.\ref{fig2}(d) it follows
\begin{eqnarray}
I(X;Y;Z)&\leq& 4H(X,Y,Z)-H(X,Y)
\nonumber\\
&&-H(X,Z)-H(Y,Z)
\label{eqn16cc}
\end{eqnarray}
and
\begin{eqnarray}
I(X;Y;Z)&>&\frac{4}{3}(H(X)+H(Y)+H(Z))
\nonumber
\\
&&-H(X,Y)-H(X,Z)-H(Y,Z)
\label{eqn16dd}
\end{eqnarray}

\textit{Example 1}. Consider a Greenberger-Horne-Zeilinger (GHZ)-type distribution:
\begin{eqnarray}
P_{ghz}=a[000]+(1-a)[111]
\end{eqnarray}
where $a\in [0,1]$. This distribution can be generated by local projection measurements under the computation basis $\{|0\rangle, |1\rangle\}$ on a GHZ state \cite{GHZ}: $\sqrt{a}|000\rangle+\sqrt{1-a}|111\rangle$. It is easy to prove that $I(X;Y;Z)\geq0$ for $a\geq0$. Moreover, we show that this distribution is generated from a triangle network in Fig.\ref{fig2}(d) by violating the inequality (\ref{eqn16b}) for $a>0$.

\textit{Example 2}. Consider a mixture of GHZ-type distribution and W-type distribution as
\begin{eqnarray}
P_{xyz}=p P_{ghz}+(1-p)P_{w}
\label{gw}
\end{eqnarray}
where $P_{ghz}=\frac{1}{2}[000]+\frac{1}{2}[111]$, $P_{w}=\frac{1}{3}[001]+\frac{1}{3}[010]+\frac{1}{3}[100]$, and $p\in [0,1]$. Here, the W-type distribution $P_{w}$ can be generated by local projection measurements under the computation basis $\{|0\rangle, |1\rangle\}$ on a W state \cite{Dur}: $\frac{1}{\sqrt{3}}(|001\rangle+|010\rangle+|100\rangle)$. It follows from $I(X;Y;Z)\leq 0$ for $p\leq 0.814$ assisted by numeric evaluations. Moreover, the distribution is generated from a triangle network in Fig.\ref{fig2}(d) by violating the inequality (\ref{eqn16a}) or (\ref{eqn16b}) for $p\geq 0.836$.

\begin{figure}
\begin{center}
\resizebox{240pt}{180pt}{\includegraphics{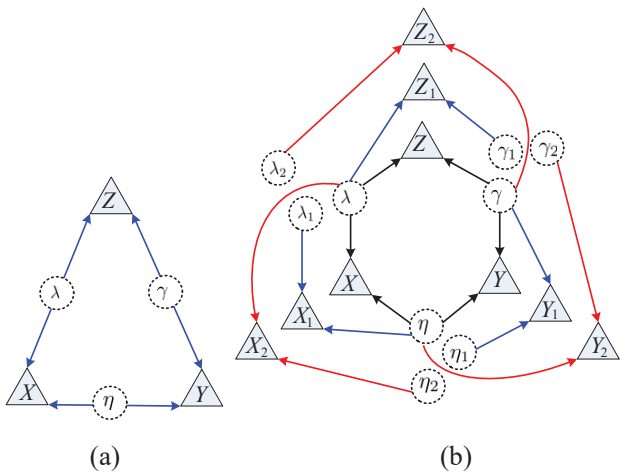}}
\end{center}
\caption{\small (Color online) (a) Triangle network consisting of three latent variables $\lambda,\gamma$ and $\eta$. (b) Second-order inflation of triangle network. Here, the network consisting of $\lambda_1,\gamma_1$ and $\eta_1$ denotes the first-order inflation while $\lambda_2,\gamma_2$ and $\eta_2$ consist of the second-order.
}
\label{fig3}
\end{figure}

\textit{Example 3}. Consider a generalized W-type distribution \cite{Dur}:
\begin{eqnarray}
P_{w}=a[001]+b[010]+(1-a-b)[100],
\end{eqnarray}
where $a,b\geq 0$ and $0\leq a+b\leq 1$. We show that this distribution implies a negative Shannon mutual information of $I(X;Y;Z)<0$ for any $a,b$ (Appendix D). Unfortunately, this distribution cannot be verified by using the inequalities (\ref{eqn16a}) and (\ref{eqn16b}). Instead, we verify any one of the following distributions (Appendix E)
\begin{eqnarray}
P_{xyz}=a[001]+b[010]+c[100]+d[011],
\label{E0a}
\\
P_{xyz}=a[001]+b[011]+c[100]+d[110],
\label{E0b}
\end{eqnarray}
and its permutations using the second-order inflation method \cite{Blou}, as shown in Fig.\ref{fig3}, where $a,b,c,d\geq0$ and $a+b+c+d=1$. Here, the triangle network consisting of $X_1,Y_1,Z_1$ and $\lambda_1,\gamma_1,\eta_1$ is the first-order inflation of the initial network consisting of $X,Y,Z$ and variables $\lambda,\gamma,\eta$. Moreover, the triangle network consisting of $X_2,Y_2,Z_2$ and $\lambda_2,\gamma_2,\eta_2$ is the second-order inflation of the initial network.

\subsection{Generic negative Shannon mutual information}

Results 1 and 2 show specific network decompositions can be arisen from of negative Shannon mutual information. Our consideration here is for exploring its converse problem, that is, what information can be learned from a given tripartite network configuration? In general, for a given tripartite Bayesian network $\mathcal{N}$ (consisting of one, two, or three independent measurable variables), define the minimal tripartite mutual information as
\begin{eqnarray}
I_{min}(X;Y;Z)=\min_{P_{xyz}}\{I(X;Y;Z)\}
\label{eqn18}
\end{eqnarray}
where the minimum is over all the possible distributions $P_{xyz}$ generated from  $\mathcal{N}$. Informally, any one in a tripartite Bayesian network can locally generate additional information for others than its absence. This shows generic negative Shannon mutual information for tripartite networks (Appendix F).

\textbf{Result 3}. Any distribution from tripartite Bayesian networks satisfies
\begin{eqnarray}
I_{min}(X;Y;Z)\leq 0
\label{eqn19}
\end{eqnarray}

Result 3 intrigues an interesting indicator of any tripartite network by defining the tripartite information increasing as
\begin{eqnarray}
\Delta
&:=&\max\{I(X;Y|Z)-I(X;Y)\}
\nonumber
\\
&=&-I_{min}(X;Y;Z)
\label{eqn20}
\end{eqnarray}
This can be used to characterize how much information can be built by local operations and classical communication (LOCC) of one party. It is of a fundamental rule of management science. One example is the organizational theory \cite{Kem}, where the indicator $\Delta$ can be used to characterize the increasing information by group behaviors associated with the network relationship. This may be extended and applied for general organizational theory beyond the scope of this paper.

\subsection{Negative Shannon mutual information from general networks}

\begin{figure}
\begin{center}
\resizebox{200pt}{95pt}{\includegraphics{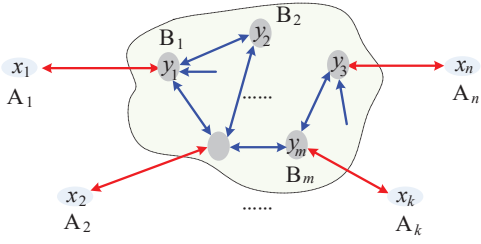}}
\end{center}
\caption{\small (Color online) Schematic $n$-independent quantum networks. Here, each pair of the nodes $\textsf{A}_1,\cdots, \textsf{A}_n$ has not shared any entanglement. While each node $\textsf{A}_i$ may share some entanglement with the node $\textsf{B}_j$.
}
\label{fig4}
\end{figure}

Results 1-3 shows the negative Shannon mutual information of three variables imply the compatible Bayesian networks. A natural problem is to explore general networks. Our method here is from so-called multipartite independent networks \cite{Luo2018}, that is, there are some nodes which have not shared any entanglement. Especially, consider an $n+m$-partite quantum network $\mathcal{N}_n$ consisting of generalized EPR states \cite{EPR}: { $|\phi_1\rangle=\cos\theta_i|00\rangle+\sin\theta_i|11\rangle$ with $\theta_i\in(0,\frac{\pi}{2})$}, $i=1, \cdots, N$. Our goal here is to consider the multipartite independent network \cite{Luo2018}. Especially, assume that there are $n$ number of nodes $\textsf{A}_1, \cdots, \textsf{A}_n$ in $\mathcal{N}_n$ with $n\geq2$ such that each pair of them has not shared any entanglement, as shown in Fig.\ref{fig4}.

Denote $X_i\in \mathcal{X}_i$ and $Y_j\in \mathcal{Y}_j$ as the respective outcomes of $\textsf{A}_i$ and $\textsf{B}_j$ under local projection measurements, $i=1, \cdots, n; j=1, \cdots, m$. Our main result  here is to prove that any general $n$-independent quantum network $\mathcal{N}_n$ (with $n\geq 3$) shows different features beyond $\mathcal{N}_2$, that is, the chain network in Fig.\ref{fig2}(b). Specially, we show that both negative and positive mutual information can be generated from $\mathcal{N}_n$ with $n\geq 3$. Here, by using Eq.(\ref{eqn2}) iteratively and the equality $H(X_1,\cdots, X_j|X_{j+1})=H(X_1,\cdots, X_j,X_{j+1})-H(X_{j+1})$ the mutual information of multivariate is defined as
\begin{eqnarray}
I(X_1;\cdots;X_n;Y)=I(X_1;\cdots;X_n)-I(X_1;\cdots;X_n|Y)
\label{mutna}
\end{eqnarray}
where $I(X_1;\cdots;X_n)$ denotes the mutual information of variables $X_1, \cdots, X_n$ and can be defined by $I(X_1;\cdots;X_n)=\sum_{\mbox{\small odd } i\leq n}\sum_{1\leq j_1<\cdots< j_i\leq n}H(X_{j_1},\cdots, X_{j_i})-\sum_{\mbox{\small even } s\leq n}\sum_{1\leq \ell_1<\cdots< \ell_i\leq n}H(X_{\ell_1},\cdots, X_{\ell_i})$, and $I(X_1;\cdots;X_n|Y)$ denotes the mutual information of $X_1, \cdots, X_n$ conditional on the outcomes of $Y=Y_1\cdots{}Y_m$ and can be defined as $I(X_1;\cdots;X_n)=\sum_{\mbox{\small odd } i\leq n}\sum_{1\leq j_1<\cdots< j_i\leq n}H(X_{j_1},\cdots, X_{j_i}|Y)-\sum_{\mbox{\small even } s\leq n}\sum_{1\leq \ell_1<\cdots< \ell_i\leq n}H(X_{\ell_1},\cdots, X_{\ell_i}|Y)$.

We firstly show some local operations may generate positive Shannon information for others as (Appendix G)
\begin{eqnarray}
&&I(X_1;\cdots;X_n)=0,
\nonumber
\\
&&I(X_1;\cdots;X_n|Y)>0,
\label{Gga1}
\end{eqnarray}
where the joint probability distribution $P_{X_1\cdots{}X_nY}$ is obtained from local measurements on the network $\mathcal{N}_k$. This implies negative Shannon mutual information as
\begin{eqnarray}
I(X_1;\cdots;X_n;Y)<0
\label{Gga2}
\end{eqnarray}
Moreover, there are some local operations may generate negative Shannon mutual information for others as (Appendix G)
\begin{eqnarray}
&&I(X_1;\cdots;X_n)=0,
\\
&&I(X_1;\cdots;X_n|Y)<0.
\label{Gga3}
\end{eqnarray}
This implies positive Shannon mutual information as
\begin{eqnarray}
I(X_1;\cdots;X_n;Y)>0,
\label{Gga4}
\end{eqnarray}
which is different from Eq.(\ref{Gga2}). Thus the general $k$-independent quantum network with $k\geq3$ can generate both negative and positive Shannon mutual information beyond Result 1 for tripartite chain network even if both have similar Bell nonlocality \cite{Luo2018}. This intrigues new kinds of non-Shannon-type information for any independent set.

\textit{Example 4}. One example is the joint distribution { in Eq.(\ref{eqn11})}. Another example is given by
\begin{eqnarray}
P_{xyz}=\tilde{\sum}_{i,j}p_iq_j[x_iy_jz_{ij}],
\label{exme1}
\end{eqnarray}
with three finite-dimensional random variables $X, Y, Z$, where $\{p_i\}$ and $\{q_j\}$ are probability distributions. Consider classical transformation $g: z_{ij}\mapsto x_iy_j$, it follows
\begin{eqnarray}
P_{xyg(z)}=\tilde{\sum}_{i,j}p_iq_j[x_iy_jx_{i}y_j]=\tilde{\sum}_ip_i[x_ix_i]\tilde{\sum}_jq_j[y_jy_j],
\end{eqnarray}
which raises two independent joint distributions of $\tilde{\sum}_ip_i[x_ix_i]$ and $\tilde{\sum}_jq_j[y_jy_j]$. It is easy to check that the joint distribution $P_{xyg(z)}$ satisfying the constriction { in Eq.(\ref{eqn9})}, i.e., a negative information of $I(X;Y;Z)<0$.  From Result 1, there is a tripartite chain network compatible with the distribution { in Eq.(\ref{exme1})}. From Eq.(\ref{Gga1}) the present example can be extended for general star network which is compatible with the joint distribution
\begin{eqnarray}
P_{x_1\cdots{}x_ny}=\tilde{\sum}_{i_1,\cdots,i_n}p_{i_1}\cdots p_{i_n}[x_{i_1}\cdots x_{i_n}y_{i_1\cdots i_n}],
\label{exme1b}
\end{eqnarray}
with $n+1$ finite-dimensional variables $X_1, \cdots, X_n, Y$, where $\{p_{i_j}\}$ are probability distributions, $j=1, \cdots, n$.

\textit{Example 5}.  For the triangle network one example is given by Eq.(\ref{eqn14}). Another is given by
\begin{eqnarray}
P_{xyz}=\tilde{\sum}_{i,j,k}p_iq_jr_k[x_{ij}y_{jk}z_{ki}]
\label{exme2}
\end{eqnarray}
with three finite-dimensional variables $X,Y,Z$, where $\{p_i\}, \{q_j\}$ and $\{r_k\}$ are probability distributions. Consider local mappings $g_1: x_{ij}\mapsto (i,j)$, $g_2: y_{jk}\mapsto (j,k)$ and $g_3: z_{k,i}\mapsto (k,i)$. It follows a new joint distribution
\begin{eqnarray}
P_{g_1(x)g_2(y)g_3(z)}=\tilde{\sum}_{i,j,k}p_iq_jr_k[ij,jk,ki]
=\tilde{\sum}_{i}p_i[ii]\tilde{\sum}_jq_j[jj]\tilde{\sum}_kr_k[kk]
\label{exme3}
\end{eqnarray}
which raises three independent joint distributions of $\tilde{\sum}_ip_i[ii]$, $\tilde{\sum}_jq_j[jj]$ and $\tilde{\sum}_kr_k[kk]$. This means the joint distribution in Eq.(\ref{exme3}) satisfy the conditions { in Eqs.(\ref{eqn13}) and (\ref{eqn15})}. From Result 2, there is a tripartite triangle network consisting of three variables compatible with the distribution { in Eq.(\ref{exme2})}.

\section{Conclusions}

For tripartite Shannon mutual information Results 1 and 2 imply a compatible chain network for special decomposition of its joint distribution in two cases. This means the negative Shannon mutual information hide different network configurations compatible with specific decompositions of joint distributions. This is different from recent results for featuring higher-order statistical correlations \cite{YSL} which takes use of Euler diagram corresponding to Shannon information. The main reason is that negative Shannon mutual information can be featured by using non-Shannon inequalities beyond Euler diagram. Here, the set bounded by Shannon-type information inequalities is denoted as polymatroidal region. A general problem is to determine whether all the polymatroids are entropic. Especially, the negative Shannon information of three random variables implies the existence of non-entropic polymatroids on the boundary \cite{Zhang1997,Matus,Yeung}. This is further extended for four or more random variables, which even allow unconstrained non-Shannon-type information inequalities \cite{Zhang1997}. A natural problem is then to explore the Markov compatibility of general non-Shannon-type information. Another problem is to explore different formations of Shannon-type information \cite{Ay}.

In summary, we provided an operational characterization of negative Shannon mutual information. The main idea is inspired by Bayesian networks. We have investigated the intrinsic network compatibility of all tripartite joint distributions. Similar results are proved for its quantum realizations. This provided a general method for experimentally verifying negative Shannon information in a device-independent manner. These results should be interesting in the information theory, deep learning, quantum nonlocality, and quantum networks.

\section*{Acknowledgements}

We are grateful to Jingyun Fan, Yali Mao and Zhengda Li for discussions. This work was supported by the National Natural Science Foundation of China (Nos.62172341,61772437), and Shenzhen Institute for Quantum Science and Engineering, Southern University of Science and Technology (Grant No. SIQSE202105).

\appendix

\section{The nonnegative tripartite Shannon mutual information}

In this section, we prove $I(X;Y;Z)\geq 0$ if three random variables $\{X,Y, Z\}$ on finite sample spaces consist of a Markov chain, that is, $X\to Y\to Z$, $X\to Z\to Y$, or $Y\to X\to Z$. The following proof holds for each case because of the symmetry of $I(X;Y;Z)$, that is, $I(X;Y;Z)=I(X;Z;Y)=I(Y;X;Z)$.

Assume that $X, Y$ and $Z$ consist of a Markov chain $X\to Y\to Z$. This implies that $X$ and $Z$ are independent conditional on $Y$. It follows that $p_{xyz}=p_{x|y}p_{z|y}$, where $P_{xyz}=\tilde{\sum}_{xyz}p_{xyz}[xyz]$ is the joint distribution of $X, Y$ and $Z$, and $p_{x|y}$ and $p_{z|y}$ are marginal conditional distributions. This implies
\begin{eqnarray}
I(X;Z|Y)&=&H(YZ)+H(XZ)-H(Z)-H(XYZ)
\nonumber
\\
&=&0.
\end{eqnarray}
Combined with the definition of $I(X;Y;Z)$ we have
\begin{eqnarray}
I(X;Y;Z)&=&I(X;Z)-I(X;Z|Y)
\nonumber
\\
&=&I(X;Z)
\nonumber
\\
&\geq&0.
\label{A01}
\end{eqnarray}
This has proved the result.

\section{Proof of Result 1}

\subsection{Classical Bayesian networks}

\begin{figure}
\begin{center}
\resizebox{240pt}{120pt}{\includegraphics{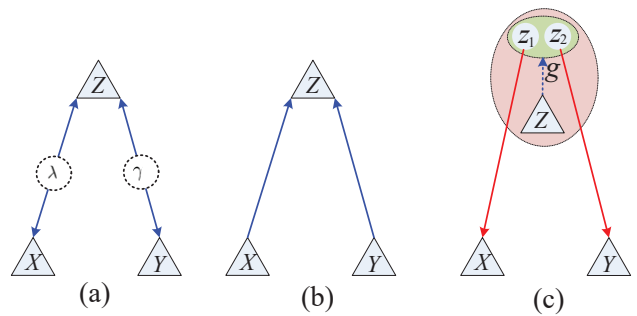}}
\end{center}
\caption{\small (Color online) Simulating a tripartite chain network. (a) Classical chain network consisting of two latent variables. (b) Chain Bayesian network for representing the distribution $P_{xyz}$ satisfying the decomposition (\ref{A2a}). Here, $g: \mathcal{Z}\to \mathcal{Z}_1\times \mathcal{Z}_2$ is classical post-processing function for any $z\in\mathcal{Z}$. (c) Chain network for representing the distribution $P_{xyz}$ satisfying the decomposition (\ref{A1b}).}
\label{figs1}
\end{figure}

The "hidden variable" (or latent variable) is firstly introduced in Ref.\cite{EPR} for arguing the incompleteness of quantum mechanics. These additional variables were from the assumptions of the causality and locality. Instead, Bohm \cite{Bohm} construct a hidden variable interpretation of elementary quantum theory. That particular interpretation has indeed a grossly nonlocal structure, according to the result to be proved by Bell \cite{Bell}. Especially, the outcome $A$ of measurement $M_A$ is depending on its physical observable and hidden variable, that is, $p(M_A)=\int_\lambda A(M_A,\lambda)\mu(\lambda)d\lambda$, where $\mu(\lambda)$ is the probability distribution of $\lambda$. Thus the expectation value of two measurements $M_A$ and $M_B$ is given by
\begin{eqnarray}
P_h(M_A,M_B)=\int d\lambda\mu(\lambda)A(M_A,\lambda)B(M_B,\lambda).
\label{AA0}
\end{eqnarray}
This is generally inconsistent with the quantum mechanical expectation value $P_q(M_A, M_B)$ on compose system $\rho_{AB}$ for more than one set of measurements, that is, $P_h(M_A, M_B)\not=P_q(M_A, M_B)$, which is firstly verified with Bell inequality \cite{Bell}. Nevertheless, they are consistent with each other for single set of measurements \cite{EPR}.

Our goal is to prove Result 1 by using classical chain network, as shown in Fig.\ref{figs1}(a). Specially, this classical realization is rigid which includes two subcases as follows:
\begin{itemize}
\item[C1.] Any distribution $P_{xyz}$ derived from local measurement on classical networks satisfies
 \begin{eqnarray}
&&I(X;Y)=0, I(X;Y|Z)\geq 0,
\label{A1a}
\end{eqnarray}
which includes negative and zero information beyond Result 1.
\item[C2.] For any distribution $P_{xyz}=\tilde{\sum}_{xyz}p_{xyz}[xyz]$ satisfying the condition in Eq.(\ref{A1a}) there exists local post-processing function $g: \mathcal{Z}\to \mathcal{Z}_1\times \mathcal{Z}_2$ satisfying
\begin{eqnarray}
P_{xyz}=P_{xyg(z_1z_2)},
\label{A1b}
\end{eqnarray}
where $P_{xyz_1z_2}$ is compatible with a chain network, that is, $P_{xyz_1z_2}=P_{xz_1}P_{yz_2}$, $P_{xz_1}=\tilde{\sum}_{xz_1}p_{xz_1}[xz_1]$ and $P_{yz_2}=\tilde{\sum}_{yz_2}p_{yz_2}[yz_2]$.
\end{itemize}

The proof is as follows.

\textbf{Case 1}. For a given chain network in Fig.\ref{figs1}(a), consider its classical realization with two independent measurable variables $\lambda_1$ and $\lambda_2$. This implies a network decomposition of any joint distribution $P_{xyz}=\tilde{\sum}_{xyz}p_{xyz}[xyz]$ as
\begin{eqnarray}
p_{xyz}=\int_{\Omega}p(x|\lambda)p(y|\gamma)p(z|\lambda,\gamma)
\mu(\lambda)\mu(\gamma)d\lambda{}d\gamma,
\label{A2a}
\end{eqnarray}
where $\Omega=\Omega_1\times \Omega_2$. It is forward to prove that
\begin{eqnarray}
p_{xy}&=&\sum_{z}p_{xyz}
\nonumber
\\
&=&\sum_z\int_{\Omega}p(x|\lambda)p(y|\gamma)p(z|\lambda,\gamma)
\mu(\lambda)\mu(\gamma)d\lambda{}d\gamma
\nonumber
\\
&=&\int_{\Omega}p(x|\lambda)p(y|\gamma)\mu(\lambda)\mu(\gamma)d\lambda{}d\gamma
\nonumber
\\
&=&\int_{\Omega_1}p(x|\lambda)\mu(\lambda)d\lambda
\int_{\Omega_2}p(y|\gamma)\mu(\gamma)d\gamma
\nonumber
\\
&=&p_xp_y,
\label{A2b}
\end{eqnarray}
where $p_x=\sum_{yz}p_{xyz}$ and $p_y=\sum_{xz}p_{xyz}$. This implies that the distribution $P_{xyz}$ defined in Eq.(\ref{A2a}) satisfies the condition in Eq.(\ref{A1a}).

\textbf{Case 2}. For a specific joint distribution $P_{xyz}=\tilde{\sum}_{x,y,z}p_{xyz}[xyz]$ with three random variables $X, Y$ and $Z$ on finite sample spaces, assume that $P_{xyz}$ satisfies $I(X;Y)=I(X;Y|Z)=0$, that is, $X$ and $Y$ are both independent and conditional independent on $Z$. In this case, we have Markovian decomposition of $p_{xyz}$ as
\begin{eqnarray}
&&p_{xyz}=p_xp_yp_{z|xy}, \forall, x,y,z,
\label{A3a}\\
&&p_{xyz}=p_zp_{x|z}p_{y|z}, \forall, x,y,z.
\label{A3b}
\end{eqnarray}
From Algorithm 1, we can obtain two chain networks: one represents the distribution $P_{xyz}$ with two edges from $X$ to $Z$ and $Y$ to $Z$, the other represents the distribution $P_{xyz}$ with two edges from $Z$ to $X$ and $Z$ to $Y$. Both imply an undirected Bayesian network. This has completed the proof.

In what follows, we prove the result by considering $P_{xyz}$ satisfies $I(X;Y)=0$ and $I(X;Y|Z)>0$. This implies a Markovian decomposition of Eq.(\ref{A3a}). From Algorithm 1, we can obtain a chain network consisting of two edges for representing the distribution $P_{xyz}$, one is from $X$ to $Z$ and the other is $Y$ to $Z$, as shown in Fig.\ref{figs1}(b), that is, two independent variables $X$ and $Y$ are the predecessors of $Z$.

We will prove the converse-directed chain Bayesian network shown in Fig.\ref{figs1}(c) for representing the distribution $P_{xyz}$ satisfying Eq.(\ref{A1b}). The distribution $P_{xyz}$ is rewritten into
\begin{eqnarray}
P_{xyz}=\tilde{\sum}_{xy}p_{xy}[xy]\times Q_{xy},
\label{A6}
\end{eqnarray}
where $p_{xy}=\sum_{z}p_{xyz}$, and $Q_{xy}$ denotes the probability distribution of $Z$ conditional on $X=x$ and $Y=y$ and is defined by
\begin{eqnarray}
Q_{xy}=\tilde{\sum}_{z}p_{z|xy}[z]
\label{A6a}
\end{eqnarray}
with $p_{z|xy}=\frac{p_{xyz}}{p_{xy}}$.

Consider a joint distribution from the chain network as
\begin{eqnarray}
P_{xyz_1z_2}&=&P_{xz_1}P_{yz_2}
\nonumber
\\
&=&\tilde{\sum}_{x=z_1,y=z_2}p_{x}p_{y}[xyz_1z_2],
\label{A7}
\end{eqnarray}
where $P_{xz_1}=\tilde{\sum}_{x,z_1}p_{xz_1}[xz_1]$ and $P_{yz_2}=\tilde{\sum}_{y,z_2}p_{yz_2}[yz_2]$. From the condition in Eq.(\ref{A1a}), it follows from Eq.(\ref{A6}) that
\begin{eqnarray}
P_{xyz}=\tilde{\sum}_{xy}p_{x}p_y[xy]\times Q_{xy}.
\label{A8}
\end{eqnarray}
Hence, it is sufficient to show that the distribution in Eq.(\ref{A8}) can be obtained from the distribution in Eq.(\ref{A7}) using local classical post-processing of $Z_1$ and $Z_2$. Note that $Q_{xy}$ in Eq.(\ref{A6a}) is a probability distribution, that is $\sum_{z}\frac{p_{xyz}}{p_{xy}}=\sum_{z}p_{z|xy}=1$. Define the following classical mapping
\begin{eqnarray}
g:[z_1=x,z_2=y] \mapsto Q_{xy}, \forall x, y,
\label{A9}
\end{eqnarray}
that is, the output of $z_1=x$ and $z_2=y$ is mapping into a probability distribution $Q_{xy}$. $g$ is a classical post-processing function of probability. In fact, define a nonnegative matrix as
\begin{eqnarray}
S_{xy}=\left(
\begin{array}{cccc}
\mathrm{g}_{1,1}(\vec{v})
\\
\vdots
\\
\mathrm{g}_{x,y}(\vec{v})
\\
\vdots
\\
\mathrm{g}_{n,n}(\vec{v})
\end{array}
\right),
\label{A10}
\end{eqnarray}
where $\vec{v}=(p_{z=1|xy},\cdots, p_{z=n|xy}, 0, \cdots, 0)$ denotes $n^2$-dimensional nonnegative vector associated with the probability distribution $Q_{xy}$, $\mathrm{g}_{i,j}$ denote $n^2$ different permutations of vector $\vec{v}$ such that $\mathrm{g}_{x,y}(\vec{v})=\vec{v}$ for each pair of $x$ and $y$. Here, we assume there are $n$ samples of $x,y$. This implies
\begin{eqnarray}
\vec{e}_{xy}\cdot{}S_{xy}=\vec{v},\forall x, y,
\label{A10a}
\end{eqnarray}
where $\vec{e}_{xy}$ denotes the $n^2$-dimensional unit vector with the $xy$-th component being $1$ and others being $0$. It is easy to prove that $S_{xy}$ are double stochastic matrices, that is, the entries are non-negative, and each row and column sums to 1. From Birkhoff's theorem \cite{Birk}, it shows that the set of doubly stochastic matrices is a convex set whose extreme points are the permutation matrices. This means that all the transformations $S_{xy}$ can be realized in statistics. Moreover, all the outputs $\{xy\}$ are distinguishable. This means that each $S_{xy}$ can be locally performed for each pair of outputs $x$ and $y$. So, we have proved that the distribution in Eq.(\ref{A8}) can be obtained from the distribution in Eq.(\ref{A7}) by using local classical post-processing of $Z_1$ and $Z_2$. From Algorithm 1, we obtain the Bayesian network in Fig.\ref{figs1}(c).

In classical realization of Bayesian network, since the variables $x$ and $y$ are independent, from Eq.(7) in the maintext there are two independent measurable variables $\lambda_1$ and $\lambda_2$ such that
\begin{eqnarray}
&&p_x=\int_{\Omega_1}p(x|\lambda)\mu(\lambda)d\lambda,
\\
&&p_y=\int_{\Omega_2}p(y|\gamma)\mu(\gamma)d\gamma,
\label{A11}
\end{eqnarray}
where $p(x|\lambda)$ and $p(y|\gamma)$ are characteristic functions, that is, $p(x|\lambda), p(y|\gamma)\in \{0, 1\}$. Note that the function in Eq.(\ref{A9}) is measurable. Combined with Eqs.(4) and (\ref{A1b}) we get from the independence of $X$ and $Y$ that
\begin{eqnarray}
p_{xyz}=\int_{\Omega}p(x|\lambda)p(y|\gamma)g(p((z_1z_2)|\lambda,\gamma))
\mu(\lambda)d\lambda\mu(\gamma)d\gamma,
\label{A12}
\end{eqnarray}
where $\Omega=\Omega_1\times \Omega_2$, as shown in Fig.\ref{figs1}(a). This has completed the proof.

\subsection{Quantum chain networks}

Our goal in this section is to prove Result 1 by using quantum chain network, as shown in Fig.2(b). Specially, the quantum realization is also rigid, that is, any distribution $P_{xyz}$ derived from local measurements on this quantum network satisfies Eq.(\ref{A1a}). Moreover, for any distribution $P_{xyz}=\tilde{\sum}_{xyz}p_{xyz}[xyz]$ satisfy the condition in Eq.(\ref{A1a}) there exists a quantum chain network consisting of two bipartite states $\rho_{AC_1}$ on Hilbert space $\mathcal{H}_{A}\otimes\mathcal{H}_{C_1}$ and $\varrho_{BC_2}$ on Hilbert space $\mathcal{H}_{B}\otimes\mathcal{H}_{C_2}$, and local measurements $\{M_x^{(A)}\}$, $\{M_y^{(B)}\}$ and $\{M_z^{(C_1C_2)}\}$ such that
\begin{eqnarray}
p_{xyz}={\rm Tr}[(M_x^{(A)}\otimes M_y^{(B)}\otimes M_z^{(C_1C_2)})\rho_{AC_1}\otimes \varrho_{BC_2}].
\label{B3}
\end{eqnarray}

In fact, consider a quantum network consisting of any bipartite states $\rho_{AC_1}$ on Hilbert space $\mathcal{H}_{A}\otimes\mathcal{H}_{C_1}$ and $\varrho_{BC_2}$ on Hilbert space $\mathcal{H}_{B}\otimes\mathcal{H}_{C_2}$. For any POVM measurements $\{M_x^{(A)}\}$, $\{M_y^{(B)}\}$ and $\{M_z^{(C_1C_2)}\}$, we have
\begin{eqnarray}
p_{xy}={\rm Tr}[(M_x^{(A)}\otimes M_y^{(B)})\rho_{A}\otimes \varrho_{B}],
\label{B4}
\end{eqnarray}
where $\rho_{A}$ and $\varrho_{B}$ are reduced density matrices of the systems $A$ and $B$, respectively. It follows that $p_{xy}=p_xp_y$ which satisfies the condition (\ref{A1a}).

In fact, the proof in Appendix B.1 above  has suggested a proof by using mixed states for quantum networks, where we can define $\rho_{AC_1}=\sum_{x,z_1}p_{xz_1}|xz_1\rangle\langle xz_1|$ and $\rho_{BC_2}=\sum_{y,z_2}p_{yz_2}|yz_2\rangle\langle yz_2|$. Each particle is measured under the computation basis $\{|x\rangle\}$, $\{|y\rangle\}$ or $\{|z_i\rangle\}$. The output of $z_1z_2$ is transformed into $z$ by using classically post-proposing function in Eq. (\ref{A9}). Note that the post-proposing function is depending of the output $z_1z_2$. This generally cannot be realized by local quantum measurements on $C_1$ and $C_2$. Instead, in what follow we consider a quantum network consisting of two entangled states. With new quantum realization, any distribution satisfying the condition in Eq.(\ref{A1a}) can be generated by local quantum measurements without classical post-progressing.

For a given distribution $P_{xyz}=\tilde{\sum}_{xyz}p_{xyz}[xyz]$ satisfying the condition in Eq. (\ref{A1a}), we get marginal distributions $P_{xy}=\tilde{\sum}_{xy}p_{xy}[xy]$, $P_{x}=\tilde{\sum}_{x}p_{x}[x]$ and $P_{y}=\tilde{\sum}_{y}p_{y}[y]$, where $p_{xy}=\sum_{z}p_{xyz}$, $p_{x}=\sum_{y,z}p_{xyz}$, and $p_{y}=\sum_{x,z}p_{xyz}$. Define a general joint state
\begin{eqnarray}
|\Phi\rangle_{ABC}=\sum_{x,y,z}\sqrt{p_{xyz}}|x,y,z\rangle
\label{B5}
\end{eqnarray}
from the joint distribution $P_{xyz}$. It follows that
\begin{eqnarray}
p_{xyz}={\rm Tr}[(|x\rangle\langle x|\otimes |y\rangle\langle y|\otimes |z\rangle\langle z|)|\Phi\rangle\langle \Phi|],
\label{B5a}
\end{eqnarray}
that is, the quantum probability under local projection measurement is compatible with the given distribution $P_{xyz}$.

From two distributions $P_{x}$ and $P_{y}$, define two entangled states as
\begin{eqnarray}
|\phi_1\rangle_{A_1C_1}=\sum_{x}\sqrt{p_{x}}|x,x\rangle,
\\
|\phi_2\rangle_{B_1C_2}=\sum_{y}\sqrt{p_{y}}|y,y\rangle.
\label{B6}
\end{eqnarray}
It follows that
\begin{eqnarray}
|\phi_1\rangle_{A_1C_1}|\phi_2\rangle_{B_1C_2}
&=&\sum_{x,y}\sqrt{p_xp_y}|x,y\rangle_{A'B'}|x,y\rangle_{C_1C_2}
\nonumber\\
&=&\sum_{x,y}\sqrt{p_{xy}}|x,y\rangle_{A'B'}|x,y\rangle_{C_1C_2}
\label{B7}
\end{eqnarray}
from the equality of $p_{xy}=p_xp_y$ because of the condition in Eq.(\ref{A1a}). The reduced density matrix of the joint system $A'B'$ is given by
\begin{eqnarray}
\rho_{A'B'}=\rho_{A'}\otimes \rho_{B'},
\label{B8}
\end{eqnarray}
where $\rho_{A'}=\sum_{x}p_x|x\rangle\langle{}x|$ and $\rho_{B'}=\sum_{y}p_y|y\rangle\langle{}y|$.

Now, from Eq.(\ref{B7}) we get the Schmidt decomposition of $|\phi_1\rangle_{A_1C_1}|\phi_2\rangle_{B_1C_2}$ as
\begin{eqnarray}
|\phi_1\rangle_{A_1C_1}|\phi_2\rangle_{B_1C_2}
=\sum_{x,y}\sqrt{p_{xy}}|x,y\rangle_{A'B'}|x,y\rangle_{C_1C_2},
\label{B9}
\end{eqnarray}
where $\{|x,y\rangle\}$ are orthogonal states from the distinguishability of samples $\{x\}$ and $\{y\}$.

From Eq.(\ref{B5}) we have
\begin{eqnarray}
|\Phi\rangle_{ABC}=\sum_{x,y}\sqrt{p_{xy}}|xy\rangle_{AB}|\varphi_{xy}\rangle_{C},
\label{B10}
\end{eqnarray}
where $|\varphi_{xy}\rangle_{C}=\sum_{z}\sqrt{\frac{p_{xyz}}{p_{xy}}}|z\rangle$ denote some states of the system $Z$. We complete the proof with two subcases, that is, $\{|\varphi_{xy}\rangle_{C}\}$ are orthogonal states or not.
\begin{itemize}
\item{} Assume that $|\varphi_{xy}\rangle_{C}$ are orthogonal states. In this case, there is a local unitary mapping given by $U: |\varphi_{xy}\rangle_{C}\mapsto |xy=z\rangle_{C_1C_2}$ because Eq.(\ref{B10}) is the Schmidt decomposition of $|\Phi\rangle$. This implies
\begin{eqnarray}
(\openone_{AB}\otimes U_C^{-1})|\phi_1\rangle_{A_1C_1}|\phi_2\rangle_{B_1C_2}=|\Phi\rangle_{ABC}.
\label{B11}
\end{eqnarray}
By performing local projection measurements under bases  $\{M_{x}=|x\rangle\langle x|_{A_1}\}$, $\{M_y=|y\rangle\langle y|_{B_1}\}$ and $\{M_{\varphi_{xy}}=|\varphi_{xy}\rangle\langle \varphi_{xy}|_{C}\}$, we get a joint distribution $P_{xyz}=\tilde{\sum}_{xyz}q_{xyz}[xyz]$ with
\begin{eqnarray}
q_{xyz}&=&{\rm Tr}[(M_x\otimes M_y\otimes M_{\varphi_{xy}})\rho_{1}\otimes \rho_2]
\nonumber
\\
&=&{\rm Tr}[(M_x\otimes M_y\otimes M_{z})
(\openone_{AB}\otimes U_C^{-1})(\rho_{1}\otimes \rho_2)]
\nonumber
\\
&=& p_{xyz}
\label{B12}
\end{eqnarray}
from Eqs.(\ref{B5a}) and (\ref{B11}), where $\rho_i=|\phi_i\rangle\langle\phi_i|$, and $\{\Pi_z=|z\rangle\langle z|_{C_1C_2}\}$. This has completed the proof.

\item{}Assume that $|\varphi_{xy}\rangle_{C}$ are not orthogonal states. In this case, we define local transformations $U_{xy}$ on the systems $C_1C_2$ and $C'$ as follows
\begin{eqnarray}
U_{xy}: |xy\rangle_{C_1C_2}|0\rangle_{C'}\mapsto |xy\rangle_{C_1C_2}|\varphi_{xy}\rangle_{C'}
\label{B13}
\end{eqnarray}
for any $x,y$, where $C'$ is an auxiliary system in the state $|0\rangle$. After these local operations being performed on the joint system $C_1C_2$, the total state of chain quantum network is changed into
\begin{eqnarray}
|\Psi\rangle_{A_1B_1C_1C_2C'}:&=&(\openone_{A_1B_1}\otimes_{xy}U_{xy})|\phi_1\rangle_{A_1C_1}|\phi_2\rangle_{B_1C_2}|0\rangle_{C'}
\nonumber
\\
&=&\sum_{x,y}\sqrt{p_{xy}}|xy\rangle_{A_1B_1}|xy\rangle_{C_1C_2}|\varphi_{xy}\rangle_{C'}.
\label{B14}
\end{eqnarray}
By performing local measurements under projections $\{M_{x}=|x\rangle\langle x|_{A_1}\}$, $\{M_y=|y\rangle\langle y|_{B_1}\}$, $\{M_{xy}=|xy\rangle\langle xy|_{C_1C_2}\}$ and $\{M_{\varphi_{xy}}=|\varphi_{xy}\rangle\langle \varphi_{xy}|_{C}\}$, we get a joint distribution $P_{xyz}=\tilde{\sum}_{xyz}q_{xyz}[xyz]$ with
\begin{eqnarray}
q_{xyz}&=&{\rm Tr}[(M_x\otimes M_y\otimes M_{xy}\otimes M_{\varphi_{xy}})\rho_1\otimes \rho_2\otimes|0\rangle\langle 0|_{C'}]
\nonumber
\\
&=&{\rm Tr}[(M_x\otimes M_y\otimes M_{xy}\otimes M_z)(\openone_{A_1B_1} \otimes_{xy}U_{xy})
\nonumber
\\
&&\times (\rho_1\otimes \rho_{2}\otimes |0\rangle\langle 0|_{C'})]
\nonumber
\\
&=& {\rm Tr}[(M_x\otimes M_y\otimes M_{xy}\otimes M_z)|\Psi\rangle\langle \Psi|]
\nonumber
\\
&=&p_{xyz}
\label{B15}
\end{eqnarray}
from Eqs.(\ref{B5a}) and (\ref{B14}). This has completed the proof.
\end{itemize}

\section{Proofs of Inequalities (19)-(23)}

For a triangle network consisting of three independent random variables $X, Y$ and $Z$ on finite sample spaces the joint distribution satisfies the following inequality \cite{Fritz} as
\begin{eqnarray}
H(X)+H(Y)+H(Z)\leq H(X,Y)+ H(Y,Z).
\label{C3}
\end{eqnarray}
Combining with the definition of $I(X;Y;Z)$ it implies that
\begin{eqnarray}
I(X;Y;Z)&=&H(X)+H(Y)+H(Z)+H(X,Y,Z)
\nonumber
\\
&&-H(X,Y)-H(Y,Z)-H(X,Z)
\nonumber
\\
&\leq & H(X,Y,Z)-H(X,Z)
\nonumber
\\
&=&H(Y|X,Z).
\label{C4}
\end{eqnarray}
Similarly, by considering the cyclic permutation of three random variables in the inequality (\ref{C3}), it follows that
\begin{eqnarray}
I(X;Y;Z)&\leq& H(X|Y,Z),
\\
I(X;Y;Z)&\leq& H(Z|X,Y).
\label{C5}
\end{eqnarray}
This implies the upper bound in the inequality (19).

Now, we prove the lower bound of triangle mutual information by using the Finner inequality \cite{Marc2019a} of
\begin{eqnarray}
p_{xyz}\leq \sqrt{p_xp_yp_z},
\label{C1}
\end{eqnarray}
where $P_{xyz}=\tilde{\sum}_{x,y,z}p_{xyz}[xyz]$ is a given joint distribution derived from triangle network consisting of three independent variables $\lambda,\gamma$ and $\eta$. $P_x=\tilde{\sum}_xp_x[x], P_y=\tilde{\sum}_yp_y[y]$ and $P_z=\tilde{\sum}_zp_z[z]$ are marginal distributions. The inequality (\ref{C1}) implies
\begin{eqnarray}
H(X,Y,Z)&=&\sum_{x,y,z}p_{xyz}\log\frac{1}{p_{xyz}}
\nonumber
\\
&\geq &\sum_{x,y,z}p_{xyz}\log\frac{1}{\sqrt{p_{x}p_yp_z}}
\nonumber
\\
&=&\frac{1}{2}\sum_{x,y,z}p_{xyz}(\log\frac{1}{p_{x}}+\log\frac{1}{p_y}+\log\frac{1}{p_z})
\nonumber
\\
&=&\frac{1}{2}(\sum_{x}p_{x}\log\frac{1}{p_{x}}+\sum_{y}p_{y}\log\frac{1}{p_y}
+\sum_{z}p_{z}\log\frac{1}{p_z})
\nonumber
\\
&=&\frac{1}{2}(H(X)+H(Y)+H(Z))
\label{C2}
\end{eqnarray}
from the increasing function of $f(x)=\log x$. Combining with the definition of $I(X;Y;Z)$, it follows the inequality (20).

For a triangle network consisting of one variable $\lambda$, its joint distribution $P_{xyz}$ satisfies as
\begin{eqnarray}
I(X;Y;Z)&\leq& 4H(X,Y,Z)-H(X,Y)
\nonumber\\
&&-H(X,Z)-H(Y,Z)
\label{eqn16c}
\end{eqnarray}
and
\begin{eqnarray}
I(X;Y;Z)&>&\frac{4}{3}(H(X)+H(Y)+H(Z))
\nonumber
\\
&&-H(X,Y)-H(X,Z)-H(Y,Z)
\label{eqn16d}
\end{eqnarray}
In fact, we have recently proved \cite{Luo2021} that
\begin{eqnarray}
p_{xyz}\leq (p_xp_yp_z)^{\frac{1}{3}}
\label{C3}
\end{eqnarray}
for any $p_{xyz}$. Similar to the inequality (\ref{C2}) we have
\begin{eqnarray}
H(X,Y,Z)&\leq &\frac{1}{3}(H(X)+H(Y)+H(Z))
\label{C4}
\end{eqnarray}
This implies the inequalities (22) and (23) from the definition of $I(X;Y;Z)$.

\section{Negative Shannon information of the W-type distribution}

\begin{figure}
\begin{center}
\resizebox{200pt}{150pt}{\includegraphics{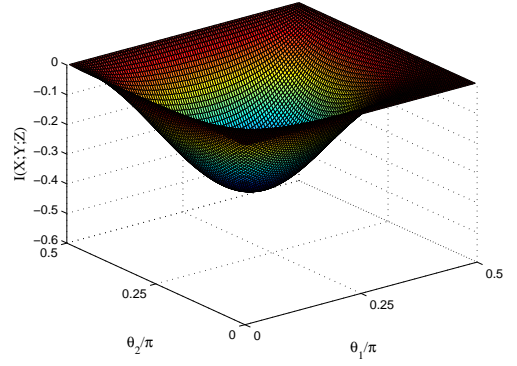}}
\end{center}
\caption{\small (Color online) The total mutual information $I_w(X;Y;Z)$ of the W-type distribution $P_{xyz}=a[001]+b[010]+c[100]$. Here, $a=\cos^2\theta_1,
b=\sin^2\theta_1\cos^2\theta_2$ and $c=\sin^2\theta_1\sin^2\theta_2$, $\theta_1,\theta_2 \in (0, \frac{\pi}{2})$.}
\label{figs2}
\end{figure}

Consider the following W-type distribution
\begin{eqnarray}
P_{xyz}=a[001]+b[010]+c[100]
\label{F1}
\end{eqnarray}
with $a+b+c=1$ and $a,b,c\geq0$. It yields to marginal distributions as
\begin{eqnarray}
P_{x}&=&(1-c)[0]+c[1]
\nonumber\\
P_y&=&(1-b)[0]+b[1]
\nonumber\\
P_z&=&(1-a)[0]+a[1]
\nonumber\\
P_{xy}&=&a[00]+b[01]+c[10]
\nonumber\\
P_{yz}&=&c[00]+a[01]+b[10]
\nonumber\\
P_{xz}&=&b[00]+a[01]+c[10]
\end{eqnarray}
From the definitions of Shannon entropy in Eq.(1) we get
\begin{eqnarray}
H(X)&=&-c\log_2c-(1-c)\log_2(1-c),
\nonumber\\
H(Y)&=&-b\log_2b-(1-b)\log_2(1-b),
\nonumber\\
H(Z)&=&-a\log_2a-(1-a)\log_2(1-a),
\nonumber\\
H(X,Y)&=&H(X,Z)=H(Y,Z)=H(X,Y,Z)
\nonumber\\
&=&-a\log_2a-b\log_2b-c\log_2c
\end{eqnarray}
This implies Shannon mutual information as
\begin{eqnarray}
I_w(X;Y;Z)&=&H(X)+H(Y)+H(Z)+H(X,Y,Z)
\nonumber
\\
&&-H(X,Y)-H(X,Z)-H(Y,Z)
\nonumber
\\
&=&a\log_2a+b\log_2b+c\log_2c
\nonumber
\\
&&-(1-a)\log_2(1-a)-(1-b)\log_2(1-b)
\nonumber\\
&&-(1-c)\log_2(1-c)
\nonumber
\\
&<&0
\end{eqnarray}
from \cite[Appendix A]{Dual}. Here, the total Shannon mutual information of the W-type distribution is shown in Fig.S\ref{figs2}.

Moreover, consider classical transformations of the W-type distribution in Eq.(\ref{F1}) with local doubly stochastic matrices defined by
\begin{eqnarray}
&&U_x=\left(
\begin{array}{ccc}
\cos^2\gamma_1 & \sin^2\gamma_1
\\
\sin^2\gamma_1 & \cos^2\gamma_1
\end{array}
\right),
\label{Fd7}
\\
&&U_y=\left(
\begin{array}{ccc}
\cos^2\gamma_2 & \sin^2\gamma_2
\\
\sin^2\gamma_2 & \cos^2\gamma_2
\end{array}
\right)
,
\label{Fd8}
\\
&&U_z=\left(
\begin{array}{ccc}
\cos^2\gamma_3 & \sin^2\gamma_3
\\
\sin^2\gamma_3 & \cos^2\gamma_3
\end{array}
\right)
\label{Fd9}
\end{eqnarray}
Define two quantities
\begin{eqnarray}
&&I_1(X;Y;Z)=\max_{\gamma_1,\gamma_2,\gamma_3}\{I(X;Y;Z)\},
\label{Fd10}
\\
&&I_2(X;Y;Z)=\min_{\gamma_1,\gamma_2,\gamma_3}\{I(X;Y;Z)\},
\label{Fd11}
\end{eqnarray}

Interestingly, assisted by numeric simulations the maximal Shannon mutual information $I_1(X;Y;Z)$ over all probabilistic transformations in Eqs.(\ref{Fd7})-(\ref{Fd9}) is almost constant for any $a, b, c$, that is, $I_1(X;Y;Z)\approx 0.5307$. Meanwhile, the W-type distribution in Eq.(\ref{F1}) achieves the minimal Shannon mutual information according to the local probabilistic transformations in Eqs.(\ref{Fd7})-(\ref{Fd9}), that is,
\begin{eqnarray}
I_w(X;Y;Z)\approx I_{2}(X;Y;Z)
 \label{Fd12}
\end{eqnarray}
up to the error scale of $10^{-7}$.

\section{Generalized W-type distributions in Eqs.(27) and (28)}

In this section we firstly prove how to verify generalized W-type distributions in Eqs.(27) and (28) by using the network inflation \cite{Blou}. And then, we prove the negative Shannon mutual information for these distributions.

\subsection{Witnessing generalized W-type distribution}

In this section we prove that any one of the following W-type distributions
\begin{eqnarray}
P_{xyz}=a[001]+b[010]+c[100]+d[011]
\label{EE0a}
\\
P_{xyz}=a[001]+b[010]+c[100]+d[101]
\label{EE0b}
\\
P_{xyz}=a[001]+b[010]+c[100]+d[110]
\label{EE0c}
\\
P_{xyz}=a[010]+b[011]+c[100]+d[101]
\label{EE0d}
\\
P_{xyz}=a[001]+b[010]+c[101]+d[110]
\label{EE0e}
\\
P_{xyz}=a[001]+b[011]+c[100]+d[110]
\label{EE0f}
\end{eqnarray}
is compatible with tripartite network consisting of one variable for any $a,b,c,d\in (0,1)$ and $a+b+c+d=1$ using the second-order inflation method \cite{Blou}. Here, we only consider the distributions in Eqs.(\ref{EE0a}) and (\ref{EE0d}) while similar proof holds for other distributions from the permutations of three variables.

\textbf{Case 1}. We prove the distribution $P_{xyz}$ in Eq.(\ref{EE0a}) which is compatible with the tripartite network consisting of three latent variables, as shown in Fig.3(a).

From Eq.(\ref{EE0a}) we get the marginal distributions as follows
\begin{eqnarray}
P_{xy}&=&a[00]+c[10]+(b+d)[01],
\label{EE1a}\\
P_{yz}&=&c[00]+b[10]+a[01]+d[11],
\\
P_{xz}&=&c[10]+b[00]+(a+d)[01],
\\
P_{x}&=&(b+a+d)[0]+c[1],
\\
P_{y}&=&(a+c)[0]+(b+d)[1],
\\
P_{z}&=&(b+c)[0]+(a+d)[1]
\label{EE1b}
\end{eqnarray}
for any $a,b,c\in(0,1)$ and $a+b+c+d=1$. Suppose that $P_{xyz}$ is compatible with the triangle network consisting of random variables $X,Y$ and $Z$ (generated from latent variables $\lambda,\gamma$ and $\eta$), as shown in Fig.3(a). A second-order spiral inflation is shown in Fig.3(b). Consider the first-order inflation network consisting of random variables $X_1,Y_1$ and $Z_1$ (generated from latent variables $\lambda_1,\gamma_1$ and $\eta_1$). From Lemma 4 \cite{Blou}, it follows all the marginal distributions of $\{ABC\}, \{XZ_1\}, \{Y_1Z\}, \{X_1Y\}, \{X_1\}, \{Y_1\} $ and $\{Z_1\}$ which have the same marginal distribution with Eqs.(\ref{EE1a})-(\ref{EE1b}) should be compatible with the first-order inflation network in Fig.3(b), that is,
\begin{eqnarray}
P_{xz_1}&=&c[10]+b[00]+(a+d)[01],
\label{EE2}\\
P_{y_1z}&=&c[00]+b[10]+a[01]+d[11],
\label{EE3}\\
P_{x_1y}&=&a[00]+c[10]+(b+d)[01],
\label{EE4}\\
P_{x_1}&=&P_x, P_{y_2}=P_y, P_{z_1}=P_z
\label{EE7}
\end{eqnarray}
From Eq.(\ref{EE2}) it implies that $X=0$ whenever $Z_1=1$. Similarly, we get $Y=0$ whenever $X_1=1$ from Eq.(\ref{EE4}), that is,
\begin{eqnarray}
{\rm Pr}[X=0|Z_1=1]&=&{\rm Pr}[Y=0|X_1=1]=1
\label{EE8}
\end{eqnarray}

Consider the second-order inflation network consisting of random variables $X_2,Y_2$ and $Z_2$ (generated from latent variables $\lambda_2,\gamma_2$ and $\eta_2$). From Lemma 4 \cite{Blou}, it follows all the marginal distributions of $\{XYZ\}, \{YZ_2\}, \{XY_2\}, \{X_2Z\}, \{X_2\}, \{Y_2\}$ and $\{Z_2\}$ which have the same marginal distribution with Eqs.(\ref{EE1a})-(\ref{EE1b}) should be compatible with the second-order inflation network in Fig.3(b), that is,
\begin{eqnarray}
P_{yz_2}&=&c[00]+b[10]+a[01]+d[11],
\label{EE12}\\
P_{xy_2}&=&a[00]+c[10]+(b+d)[01],
\label{EE13}\\
P_{x_2z}&=&c[10]+b[00]+(a+d)[01],
\label{EE14}\\
P_{x_2}&=&P_x, P_{y_2}=P_y, P_{z_2}=P_z
\label{EE17}
\end{eqnarray}
From Eq.(\ref{EE13}) it implies that $X=0$ whenever $Y_2=1$. Similarly, we get $Z=0$ whenever $X_2=1$ from Eq.(\ref{EE14}), that is,
\begin{eqnarray}
{\rm Pr}[X=0|Y_2=1]&=&{\rm Pr}[Z=0|X_2=1]=1
\label{EE18}
\end{eqnarray}

From Fig.3(b) the inflation shows the random variables $X_1, Y_1, Z_1, X_2, Y_2, Z_2$ should be marginally independent in any compatible distribution. From Eqs.(\ref{EE1b}) and (\ref{EE7}) it follows that
\begin{eqnarray}
{\rm Pr}[X_1=Z_1=X_2=Y_2=1]\not=0
\end{eqnarray}
Combined with Eqs.(\ref{EE8}) and (\ref{EE18}), it follows that
\begin{eqnarray}
{\rm Pr}[X=Y=Z=0]\not=0
\end{eqnarray}
which is contradicted to Eq.(\ref{EE0a}). This has proved that the distribution in Eq.(\ref{EE0a}) is incompatible with the tripartite network in Fig.3(a).

\textbf{Case 2}. We prove the distribution in Eq. (\ref{EE0d}) is compatible with the tripartite network consisting of three latent variables, as shown in Fig.3(a).

From Eq.(\ref{EE0d}) we get the marginal distributions as follows
\begin{eqnarray}
P_{xy}&=&(a+b)[01]+(c+d)[10],
\label{EEE1a}\\
P_{yz}&=&c[00]+a[01]+d[10]+b[11],
\\
P_{xz}&=&a[00]+b[01]+c[10]+d[11],
\\
P_{x}&=&(a+b)[0]+(c+d)[1],
\\
P_{y}&=&(c+d)[0]+(a+b)[1],
\\
P_{z}&=&(a+c)[0]+(b+d)[1]
\label{EEE1b}
\end{eqnarray}
for any $a,b,c,d\in(0,1)$ and $a+b+c+d=1$. Suppose that $P_{xyz}$ is compatible with the triangle network consisting of three latent variables, as shown in Fig.3(a). From Lemma 4 \cite{Blou} it follows all the marginal distributions of sets $\{ABC\}, \{XZ_1\}, \{Y_1Z\}, \{X_1Y\}, \{X_1\}, \{Y_1\}$ and $\{Z_1\}$ which have the same marginal distribution with Eqs.(\ref{EEE1a})-(\ref{EEE1b}) should be compatible with the first-order inflation network in Fig.3(b), that is,
\begin{eqnarray}
P_{y_1z}&=&c[00]+a[01]+d[10]+b[11],
\label{EEE2}\\
P_{x_1y}&=&(a+b)[01]+(c+d)[10],
\label{EEE3}\\
P_{xz_1}&=&a[00]+b[01]+c[10]+d[11],
\label{EEE4}\\
P_{x_1}&=&P_x, P_{y_1}=P_y, P_{z_1}=P_z.
\label{EEE7}
\end{eqnarray}
From Eq.(\ref{EEE3}) it implies that $Y=0$ whenever $X_1=1$, that is,
\begin{eqnarray}
{\rm Pr}[Y=0|X_1=1]=1.
\label{EEE8}
\end{eqnarray}

Moreover, all the marginal distributions of $\{XYZ\}, \{YZ_2\}, \{XY_2\}, \{X_2Z\}, \{X_2\}, \{Y_2\}$, and $\{Z_2\}$ which have the same marginal distribution with Eqs.(\ref{EEE1a})-(\ref{EEE1b}) should be compatible with the second-order inflation network in Fig.3(b), that is,
\begin{eqnarray}
P_{yz_2}&=&c[00]+a[01]+d[10]+b[11],
\label{EEE12}\\
P_{xy_2}&=&(a+b)[01]+(c+d)[10],
\label{EEE13}\\
P_{x_2z}&=&a[00]+b[01]+c[10]+d[11],
\label{EEE14}\\
P_{x_2}&=&P_x, P_{y_2}=P_y, P_{z_2}=P_z.
\label{EEE17}
\end{eqnarray}
From Eq.(\ref{EEE13}) it implies that $X=0$ whenever $Y_2=1$, that is,
\begin{eqnarray}
{\rm Pr}[X=0|Y_2=1]=1
\label{EEE18}
\end{eqnarray}
From Eqs.(\ref{EEE1b}) and (\ref{EEE7}) it follows that
\begin{eqnarray}
{\rm Pr}[X_1=Z_2=1]\not=0.
\end{eqnarray}
Combined with Eqs.(\ref{EEE8}) and (\ref{EEE18}), it follows that
\begin{eqnarray}
{\rm Pr}[X=Y=0]\not=0,
\end{eqnarray}
which is contradicted to Eq.(\ref{EE0d}). This has proved that the distribution in Eq.(\ref{EE0d}) is incompatible with the tripartite network in Fig.3(a).

\subsection{Negative Shannon mutual information}

Here, we firstly consider the W-type distribution in Eq.(\ref{EE0a}). We show that its Shannon mutual information satisfies $I_{ws}(X;Y;Z)\leq 0$ for the subcase of $a=c$ or $b=c$. Similar proofs hold for other two distributions in Eqs.(\ref{EE0b}) and (\ref{EE0c}). In fact, from the marginal distributions in Eqs.(\ref{EE1a})-(\ref{EE1b}) we get
\begin{eqnarray}
H(X)&=&-c\log_2c-(1-c)\log_2(1-c),
\nonumber\\
H(Y)&=&-(b+d)\log_2b-(a+c)\log_2(a+c),
\nonumber\\
H(Z)&=&-(a+d)\log_2(a+d)-(b+c)\log_2(b+c),
\nonumber\\
H(X,Y)&=&-a\log_2a-(b+d)\log_2(b+d)-c\log_2c,
\nonumber\\
H(X,Z)&=&-b\log_2b-(a+d)\log_2(a+d)-c\log_2c,
\nonumber\\
H(Y,Z)&=&H(X,Y,Z)=-a\log_2a-b\log_2b
\nonumber\\
&&-c\log_2c-d\log_2d.
\end{eqnarray}
This implies the Shannon mutual information as
\begin{eqnarray}
I_{ws}(X;Y;Z)&=&H(X)+H(Y)+H(Z)+H(X,Y,Z)
\nonumber
\\
&&-H(X,Y)-H(X,Z)-H(Y,Z)
\nonumber
\\
&=&a\log_2a+b\log_2b+c\log_2c
\nonumber\\
&&-(1-c)\log_2(1-c)-(a+c)\log_2(a+c)
\nonumber\\
&&-(b+c)\log_2(b+c).
\label{G11}
\end{eqnarray}

Now, consider the subcase of $a=c$. From Eq.(\ref{G11}) it follows that
\begin{eqnarray}
I_{ws}(X;Y;Z)&=&b\log_2(b)-(1-a)\log_2(1-a)
-2a
\nonumber\\
&&-(b+a)\log_2(b+a).
\label{G12}
\end{eqnarray}
The first-order of partial derivatives are given by
\begin{eqnarray}
\frac{\partial I_{ws}(X;Y;Z)}{\partial a}&=&\log_2b-\log_2(a+b)
\nonumber
\\
&<&0,
\\
\frac{\partial I_{ws}(X;Y;Z)}{\partial b}&=&\log_2(1-a)-\log_2(a+b)-2
\nonumber
\\
&<&0
\end{eqnarray}
for any $a, b\geq 0$. This implies that $I_{ws}(X;Y;Z)\leq 0$ from $I_{ws}(X;Y;Z)=0$ with $a=b=0$. From the symmetry of $a$ and $b$ in Eq.(\ref{G11}), we can prove the result for the subcase of $b=c$.

\begin{figure}
\begin{center}
\resizebox{200pt}{120pt}{\includegraphics{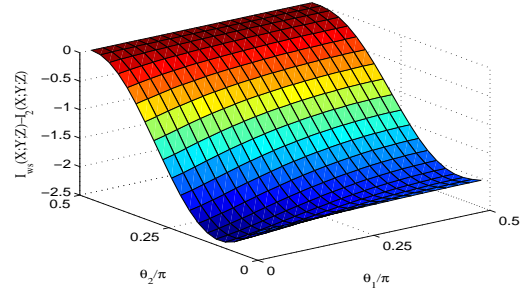}}
\end{center}
\caption{\small (Color online) The numeric difference of $I_w(X;Y;Z)$ and $I_2(X;Y;Z)$ for the W-type distribution $P_{xyz}=a[001]+b[010]+a[100]+d[011]$. Here, $a=\frac{1}{2}\cos^2\theta_1, b=\sin^2\theta_1\cos^2\theta_2$ and $d=\sin^2\theta_1\sin^2\theta_2$, $\theta_1,\theta_2 \in (0, \frac{\pi}{2})$.}
\label{figs3}
\end{figure}

Moreover, under the probabilistic transformations in Eqs.(\ref{Fd7})-(\ref{Fd9}) the maximal Shannon mutual information satisfies $I_1(X;Y;Z)\geq 0.5294$ for any $a,b$. Different from the W-type distribution in Eq.(\ref{F1}) the probabilistic transformations in Eqs.(\ref{Fd7})-(\ref{Fd9}) can change the minimal Shannon mutual information $I_2(X;Y;Z)$. Specially, we show that
\begin{eqnarray}
I_{ws}(X;Y;Z)\preceq I_2(X;Y;Z)
\label{G13}
\end{eqnarray}
for any $0\leq a,b\leq 1$. This shows a different feature of the W-type distribution in Eq. (\ref{EE0a}) from in Eq.(\ref{F1}).

Now, we consider the W-type distribution in Eq.(\ref{EE0d}). Different from the distributions in Eqs.(\ref{EE0a})-(\ref{EE0c}) its Shannon mutual information satisfies $I_{ws}(X;Y;Z)\geq 0$ for any $a,b,c,d$. In fact, from the marginal distributions in Eqs.(\ref{EEE1a})-(\ref{EEE1b}) we get
\begin{eqnarray}
H(X)&=&H(Y)=H(X,Y)
\nonumber\\
&=&-(a+b)\log_2(a+b)\nonumber\\
&&-(c+d)\log_2(c+d),
\\
H(Z)&=&-(a+c)\log_2(a+c)\nonumber\\
&&-(b+d)\log_2(b+d),
\\
H(X,Z)&=&H(Y,Z)=H(X,Y,Z)
\nonumber\\
&=&-a\log_2a-b\log_2b
\nonumber\\
&&-c\log_2c-d\log_2d.
\end{eqnarray}
This implies the Shannon mutual information satisfying
\begin{eqnarray}
I_{ws}(X;Y;Z)&=&H(X)+H(Z)-H(X,Z)
\nonumber
\\
&\geq& 0.
\label{G15}
\end{eqnarray}
Similar proofs hold for other two distributions in Eqs.(\ref{EE0e}) and (\ref{EE0f}).

\section{Proof of Result 3}

The proof of Result 3 is completed using quantum realization of Bayesian network, which includes any classical realization. We firstly prove the result for a tripartite quantum network consisting of any pure three-qubit state $|\Phi\rangle_{ABC}$ on Hilbert space $\mathcal{H}_A\otimes \mathcal{H}_B\otimes \mathcal{H}_C$. Consider its Schmidt decomposition with respect to the bipartition $A$ and $BC$ as
\begin{eqnarray}
|\Phi\rangle=\sum_{i=1}^s\lambda_i|\phi_i\rangle_{A}|\psi_i\rangle_{BC},
\label{E1}
\end{eqnarray}
where $\lambda_i$'s are Schmidt coefficients satisfying $\sum_{i}\lambda_i^2=1$, and $\{|\phi_i\rangle\}$ are orthogonal states of the particle $A$ while $\{|\psi_i\rangle\}$ are orthogonal states of joint system $BC$.

If $s=1$, $|\Phi\rangle$ is a product state. In this case, for any local POVM operators $\{M_{x}\},\{N_y\},\{L_z\}$, we have
\begin{eqnarray}
p_{xyz}&=&{\rm Tr}[|\Phi\rangle\langle \Phi|(M_{x}\otimes{}N_y\otimes{}L_z)]
\nonumber
\\
&=&{\rm Tr}(|\phi_1\rangle\langle \phi_1|M_{x})
{\rm Tr}(|\psi_1\rangle\langle \psi_1|(N_y\otimes{}L_z))
\nonumber
\\
&=&p_xp_{yz}
\label{E2}
\end{eqnarray}
for any $x,y,z$. This implies that
\begin{eqnarray}
I(X;Y;Z)=I(Y;Z)-I(Y;Z|X)=0.
\label{E3}
\end{eqnarray}
It follows that $I_{min}(X;Y;Z)=0$.

If $s=2$, $|\Phi\rangle$ is an entanglement. In this case, by using proper local unitary transformations for each particle, we obtain three entangled states as
\begin{eqnarray}
|\Phi_1\rangle&=&\lambda_1|000\rangle+\lambda_2|111\rangle,
\label{E4}
\\
|\Phi_2\rangle&=&\lambda_1|000\rangle+\lambda_2|1\rangle(\gamma_1|01\rangle
\nonumber
\\
&&+\gamma_2|10\rangle+\gamma_3|11\rangle),
\label{E5}
\\
|\Phi_3\rangle&=&\lambda_1|0\rangle(\alpha|00\rangle+\beta|11\rangle)
+\lambda_2|1\rangle(\gamma\beta|00\rangle
\nonumber
\\
&&-\gamma\alpha|11\rangle+\gamma_1|01\rangle+\gamma_2|10\rangle).
\label{E6}
\end{eqnarray}

For the GHZ-type state in Eq.(\ref{E4}) each party performs the projection measurement under the computation basis $\{|0\rangle, |1\rangle\}$. They get the tripartite probability distribution as
\begin{eqnarray}
P_{xyz}=\lambda_1^2[000]+\lambda_1^2[111],
\label{E7}
\end{eqnarray}
which can be classically transformed into
\begin{eqnarray}
P_{xyz}'=\sum_{x,y,z=0,1}\frac{1}{8}[xyz]
\label{E8}
\end{eqnarray}
by using a post-processing transformation
\begin{eqnarray}
\mathcal{T}: [i]\mapsto \frac{1}{2}[0]+\frac{1}{2}[1], i=0,1
\label{Tmap}
\end{eqnarray}
for each party. For the distribution $P_{xyz}'$ in Eq.(\ref{E8}) we get
\begin{eqnarray}
I(X;Y;Z)=0.
\label{E9}
\end{eqnarray}
This implies that $I_{min}(X;Y;Z)\leq 0$. Moreover, we can show that there are local POVM measurements such that $I_{min}(X;Y;Z)<0$ for any $\lambda_1\in (0,1)$ assisted by numeric evaluations.

For the entanglement in Eq. (\ref{E5}), each party performs the projection measurement under the computation basis $\{|0\rangle, |1\rangle\}$. They get a joint probability distribution as
\begin{eqnarray}
P_{xyz}&=&\lambda_1^2[000]+\lambda_2^2\gamma_1^2[101]+\lambda_2^2
\gamma_2^2[110]
\nonumber
\\
&&+\lambda_2^2\gamma_3^2[111],
\label{E10}
\end{eqnarray}
which can be classically transformed into the distribution in Eq.(\ref{E8}) by using the post-processing transformation in Eq. (\ref{Tmap}) for each party. This implies that $I_{min}(X;Y;Z)\leq 0$. Moreover, we can prove there are local measurements such that $I_{min}(X;Y;Z)<0$ for almost all $\lambda_i,\gamma_j\in (0,1)$ assisted by numeric evaluations.

For the entanglement in Eq.(\ref{E6}) each party performs the projection measurement under the computation basis $\{|0\rangle, |1\rangle\}$. And then, they can transform the joint probability distribution into the distribution Eq.(\ref{E8}) by using the post-processing transformation (\ref{Tmap}). This implies that $I_{min}(X;Y;Z)\leq 0$. Moreover, there are local measurements such that $I_{min}(X;Y;Z)<0$ for almost all $\lambda_i,\alpha,\beta, \gamma_j,\in (0,1)$ assisted by numeric evaluations.

Similarly result holds for any high-dimensional entanglement $|\Phi\rangle$ on Hilbert space $\mathcal{H}_A\otimes \mathcal{H}_B\otimes \mathcal{H}_C$. Here, each joint distribution under local projection measurements can be changed into a separable distribution by using the post-processing transformation in Eq.(\ref{Tmap}) for any outcome.

\section{Shannon mutual information from general networks}

Consider a general $n+m$-partite quantum network $\mathcal{N}_n$, as shown in Fig.4. Our goal here is to prove both negative and positive mutual information can be generated from $\mathcal{N}_n$ with $n\geq 3$.

\subsection{Generating positive mutual information}

In this subsection, we show some local operations may build  positive Shannon mutual information for others as
\begin{eqnarray}
&&I(X_1;\cdots;X_n)=0
\nonumber
\\
&&I(X_1;\cdots;X_n|Y)>0
\label{Gg1}
\end{eqnarray}
where the joint probability distribution $P_{X_1\cdots{}X_nY}$ is obtained from local measurements on the network $\mathcal{N}_k$. This implies negative Shannon mutual information as
\begin{eqnarray}
I(X_1;\cdots;X_n;Y)&:=&I(X_1;\cdots;X_n)
\nonumber
\\
&&-I(X_1;\cdots;X_n|Y)
\nonumber
\\
&\leq& 0,
\label{Gg1a}
\end{eqnarray}
where all the random variables $Y:=Y_1\cdots{}Y_m$ are regarded as one high-dimensional variable.

In fact, define local measurements of $\textsf{A}_i$ as $\{M_{x_i}=|x_i\rangle\langle x_i|\}$ which satisfy $\sum_{x_i}M_{x_i}=\openone_{A_i}$ with the identity operator $\openone_{A_i}$ on the state space of $\textsf{A}_i$, $i=1, \cdots, n$. Let $\{M_{y_j}\}$ with $M_{y_j}\in \{ |i\rangle\pm |2^{s_j}-i-1\rangle, i=0, \cdots, 2^{s_j-1}\}$ be local measurements of $\textsf{B}_j$ and satisfy $\sum_{y_j}M_{y_j}=\openone_{B_j}$ with the identity operator $\openone_{B_j}$ on the state space of $\textsf{B}_j$, where $s_j$ denotes the number of particle contained in the node $\textsf{B}_j$, $j=1, \cdots, n$. Since all the nodes $\textsf{A}_1, \cdots, \textsf{A}_n$ are independent, the joint probability of the outcome $x_1\cdots{}x_n$ is given by
\begin{eqnarray}
p_{x_1\cdots{}x_n}&=&{\rm Tr}(\otimes_{i=1}^nM_{x_i}\rho_{A_1\cdots{}A_n})
\nonumber
\\
&=&\otimes_{i=1}^n{\rm Tr}(M_{x_i}\rho_{A_i})
\nonumber
\\
&=&\prod_{i=1}^np_{x_i},
\label{Gg2}
\end{eqnarray}
where $\rho_{A_1\cdots{}A_n}$ denotes the reduced density matrix of $\textsf{A}_1,\cdots, \textsf{A}_n$, $\rho_{A_i}$ denotes the reduced density matrix of $\textsf{A}_i$, and $p_{x_i}$ denotes the probability of the outcome $x_i$ by $\textsf{A}_i$. From Eq.(\ref{Gg2}) it follows that
\begin{eqnarray}
I(X_1;\cdots;X_n)=0.
\label{Gg3}
\end{eqnarray}

Now, consider the projection measurement $\{M_{y_j}\}$ of $\textsf{B}_1, \cdots, \textsf{B}_m$,  and send out the measurement outcomes for $\textsf{A}_1, \cdots, \textsf{A}_n$. It is easy to check that the joint state of $\textsf{A}_1,\cdots, \textsf{A}_n$ is changed into an $n$-qubit GHZ state
\begin{eqnarray}
|\Phi\rangle_{A_1\cdots{}A_n}=\cos\varphi|0\rangle^{\otimes n}+\sin\varphi|1\rangle^{\otimes n}
\label{Gg4}
\end{eqnarray}
under proper local unitary operations \cite{Entswap1993}, where $\varphi$ depends on parameters $\theta_1,\cdots,\theta_n$. Now, under the local projection measurement $\{M_{x_i}\}$, it follows the joint distribution
\begin{eqnarray}
P_{x_1\cdots{}x_n}=\cos^2\varphi[0\cdots{}0]+\sin^2\varphi[1\cdots{}1].
\label{Gg5}
\end{eqnarray}
We obtain that
\begin{eqnarray}
I(X_1;\cdots;X_n|Y)&=&-\cos^2\varphi\log(\cos^2\varphi)
\nonumber\\
&&-
\sin^2\varphi\log(\sin^2\varphi)
\nonumber
\\
&>&0
\label{Gg6}
\end{eqnarray}
for any $\theta_1,\cdots,\theta_n\in(0,\frac{\pi}{2})$. From Eqs.(\ref{Gg3}) and (\ref{Gg6}) we get that
\begin{eqnarray}
I(X_1;\cdots;X_n;Y)<0.
\label{Gg7}
\end{eqnarray}
This has proved the result. One example is long chain network or star network \cite{Luo2018}.

\subsection{Generating positive Shannon mutual information}

In this subsection, we show some local operations may generate negative Shannon mutual information for others as
\begin{eqnarray}
&&I(X_1;\cdots;X_n)=0
\\
&&I(X_1;\cdots;X_n|Y)<0
\label{Gg8}
\end{eqnarray}
where the joint probability distribution $P_{X_1\cdots{}X_nY_1\cdots{}Y_m}$ is obtained from local measurements on the network $\mathcal{N}_k$. This implies positive Shannon mutual information as
\begin{eqnarray}
I(X_1;\cdots;X_n;Y)&=&I(X_1;\cdots;X_n)
\nonumber
\\
&&-I(X_1;\cdots;X_n|Y)
\nonumber
\\
&>& 0
\label{Gg8a}
\end{eqnarray}
which is different from Eq.(\ref{Gg1a}). This means that the $n$-independent network shows different correlations beyond the chain network in Fig.2(b) for any $n\geq 3$.

The proof is constructed as follows. Consider a special case of $\mathcal{N}_q$ as $n=3$ and $m=1$. Here, each pair of $\textsf{A}_{i}$ and $\textsf{B}$ shares one EPR state $|\phi_i\rangle$. In this case, the total state of $\mathcal{N}_q$ is given by
\begin{eqnarray}
|\Phi\rangle=\sum_{i_1i_2i_2=0,1}a_{i_1i_2i_3}|i_1i_2i_3\rangle_{A_1A_2A_3}|i_1i_2i_3\rangle_{B}
\label{Gg9}
\end{eqnarray}
where $a_{i_1i_2i_3}=\prod_{j=1}^3\cos^{i_j+1\!\!\mod\! 2}\theta_j\sin^{i_j}\theta_j$. $\textsf{B}$ performs a local measurement under the quantum Fourier basis $\{\Pi_y\}$ with $\Pi_y\in \{|\varphi_i\rangle\langle\varphi_i\}$, where $|\varphi_i\rangle$ are defined as
\begin{eqnarray}
|\varphi_i\rangle&=&\frac{1}{2}(|001\rangle+\omega^{i}|010\rangle
\nonumber
\\
&&+\omega^{2i}|100\rangle+\omega^{3i}|011\rangle),
\nonumber\\
|\varphi_{4+i}\rangle&=&\frac{1}{2}(|110\rangle+\omega^{i}|101\rangle
\nonumber
\\
&&+\omega^{2i}|000\rangle+
\omega^{3i}|111\rangle
\label{Gg10}
\end{eqnarray}
with an unit root $\omega=\sqrt{-1}$, $i=1, \cdots, 4$. After the local measurement, by proper local operation of $\textsf{A}_1$, the joint state of $\textsf{A}_1,\textsf{A}_2$ and $\textsf{A}_3$ is changed into
 \begin{eqnarray}
|\Phi_i\rangle&=&a_{001}|001\rangle+\omega^{-i}a_{010}|010\rangle
\nonumber
\\
&&+\omega^{-2i}a_{100}
|100\rangle+\omega^{-3i}a_{011}|011\rangle),
\nonumber\\
|\Phi_{4+i}\rangle&=&a_{110}|010\rangle+\omega^{-i}a_{101}|001\rangle
\nonumber
\\
&&+\omega^{-2i}a_{000}|100\rangle+\omega^{-3i}a_{111}|011\rangle,
\label{Gg11}
\end{eqnarray}
where $i=1, \cdots, 4$. Under the local projection of all parties $\textsf{A}_j$'s, it implies a joint distribution as
\begin{eqnarray}
P_{x_1x_2x_3}&=&a^2_{001}[001]+a^2_{010}[010]
\nonumber
\\
&&+a_{100}^2[100]+a^2_{011}[011],
\label{Gg12}
\end{eqnarray}
or
\begin{eqnarray}
P_{x_1x_2x_3}&=&a^2_{110}[010]+a^2_{101}[001]
\nonumber
\\
&&+a^2_{000}[100]+a^2_{111}[011]
\label{Gg13}
\end{eqnarray}
Combined with Appendix E.2, it follows that
\begin{eqnarray}
I(X_1;X_2;X_3|Y)<0
\label{Gg14}
\end{eqnarray}
for $\theta_1=\theta_3$. This has completed the proof from Eqs.(\ref{Gg3}) and (\ref{Gg14}).


\begin{thebibliography}{99}
\bibitem{Shannon}C. E. Shannon, A mathematical theory of communication, \textit{Bell Syst. Tech. J.} \textbf{27}, 379-423, 623-656 (1948).

\bibitem{Neumann}J. von Neumann, Thermodynamik quantummechanischer Gesamheiten, \textit{Gott. Nach.} \textbf{1}, 273-291(1927).

\bibitem{Jaynes1957}E. T. Jaynes, Information theory and statistical mechanics Part I. \textit{Phys. Rev.} \textbf{106}, 620-630; ibid, \textbf{107}, 171-190 (1957).

\bibitem{Honerkamp1998}J. Honerkamp, \textit{Statistical Physics}, Springer, Berlin, 1998.

\bibitem{Fuji}S. Fujishige, Polymatroidal dependence structure of a set of random variables, \textit{Info. Contr.} \textbf{39}, 55-72 (1978).


\bibitem{Zhang1997}Z. Zhang and R. W. Yeung. A non Shannon-type conditional inequality of information quantities, \textit{IEEE Trans. Inf. Theory} \textbf{43}, 1982-1986, 1997.

\bibitem{Pearl}J. Pearl,  \textit{Causality, Models, Reasoning, and Inference}, 2nd edition, Cambridge University Press, 2009.


\bibitem{EPR}A. Einstein, B. Podolsky, N. Rosen, Can quantum mechanical description of physical reality be considered complete? \textit{Phys. Rev.} \textbf{47}, 777-780 (1935).


\bibitem{NC}M. A. Nielsen and I. L. Chuang, \textit{Quantum Computation and Quantum Information}, Cambridge University Press, 10th Anniversary Edition, 2010.

\bibitem{Bell}J. S. Bell, On the Einstein-Podolsky-Rosen paradox, \textit{Phys.} \textbf{1}, 195 (1964).


\bibitem{Entswap1993}M. Zukowski, A. Zeilinger, M. A. Horne, and  A. K. Ekert, "Event-ready-detectors" Bell experiment via entanglement swapping, \textit{Phys. Rev. Lett}. \textbf{71}, 4287 (1993).

\bibitem{Cava}D. Cavalcanti, M. L. Almeida, V. Scarani \& A. Ac\'{i}n, Quantum networks reveal quantum nonlocality. \textit{Nature Commun.} \textbf{2}, 184 (2011).

\bibitem{15} C. Branciard, N. Gisin and   S. Pironio, Characterizing the nonlocal correlations created via entanglement swapping. \textit{Phys. Rev. Lett}. \textbf{104}, 170401 (2010).

\bibitem{Markov} W. Feller, \textit{An Introduction to Probability Theory and Its Applications}, Vol. 1, J. Wiley \& Sons, 3rd dition, New York, 1968.


\bibitem{SG}P. Spirtes, N. Glymour, and R. Scheienes, \textit{Causation, Prediction, and Search}, 2nd ed. MIT Press, 2001.

\bibitem{TP}J. Tian and J. Pearl. On the testable implications of causal models with hidden variables, In A. Darwiche and N. Friedman (Eds.), \textit{Uncertainty in Artificial Intelligence, Proceedings of the Eighteenth Conference}, Morgan Kaufmann: San Francisco, CA, pp.519--527, 2002.

\bibitem{KT}C. Kang and J. Tian, Inequality constraints in causal models with hidden variables, In \textit{Proceedings of the Seventeenth Annual Conference on Uncertainty in Artificial Intelligence} (UAI-06), MIT, Cambridge, MA, pp.233--240, 2006.

\bibitem{Birk}G. Birkhoff, Tres observaciones sobre el algebra lineal, \textit{Univ. Nac. Tucuman Rev. Ser. A} \textbf{5}, 147-150(1946).

\bibitem{HHH}R. Horodecki,  P.  Horodecki,  M. Horodecki,  and     K. Horodecki, Quantum entanglement, \textit{Rev. Mod. Phys}. \textbf{81}, 865 (2009).


\bibitem{Marc2019}M.-O. Renou, E. B\"{a}umer, S. Boreiri, N. Brunner, N. Gisin, and S. Beigi, Genuine quantum nonlocality in the triangle network, \textit{Phys. Rev. Lett.} \textbf{123}, 140401 (2019).


\bibitem{Spirt}P. Spirtes, R. Scheines, C. Meek, T. Richardson, C. Glymour, H. Hoijtink and A. Boomsma, \textit{TETRAD 3: Tools for Causal Modeling, Program},  Psychology Press, 1996.

\bibitem{Dur}W. D\"{u}r, G. Vidal, and  J. I. Cirac, Three qubits can be entangled in two inequivalent ways, \textit{Phys. Rev. A} \textbf{62}, 062314 (2000).


\bibitem{Blou}E. Wolfe, R. W. Spekkens, and T. Fritz, The inflation technique for causal inference with latent variables, \textit{J. Causal Infer.} \textbf{7}, 20170020 (2019).

\bibitem{NW}M. Navascu\'{e}s and E. Wolfe, The inflation technique completely solves the causal compatibility problem, \textit{J. Causal Infer.} \textbf{8}, 70-91(2020).

\bibitem{GHZ}D. M. Greenberger, M. A. Horne, and A. Zeilinger, \textit{in Bell's Theorem, Quantum Theory and Conceptions of the Universe}, edited by M. Kafatos (Kluwer, Dordrecht, 1989), pp. 69-72.

\bibitem{Kem}J. L. Kmetz, \textit{The Information Processing Theory of Organization Managing Technology Accession in Complex Systems}, 1st Edition, Ashgate Pub Ltd, 1998.

\bibitem{Luo2018} M.-X. Luo, Computationally efficient nonlinear Bell inequalities for quantum networks, \textit{Phys. Rev. Lett.} \textbf{120}, 140402 (2018).

\bibitem{Yeung}R. Yeung,  A framework for linear information inequalities, \textit{IEEE Trans. Inf. Theory} \textbf{43}, 1924-1934 (1997).

\bibitem{Matus}F. Matus, Piecewise linear conditional information inequality, \textit{IEEE Trans. Inf. Theory} \textbf{44}, 236-238 (2006).

\bibitem{Ay}N. Ay, A refinement of the common cause principle, \textit{Discrete Appl. Math.} 157, 2439-2457 (2009).

\bibitem{Bohm}D. Bohm, A Suggested Interpretation of the Quantum Theory in "Hidden" Variables. I, \textit{Phys. Rev.} \textbf{85}, 166 (1952).

\bibitem{Fritz}T. Fritzm Beyond Bell's theorem I: correlation scenarios. \textit{New J. Phys.} \textbf{14}, 103001(2012).

\bibitem{Marc2019a}M.-O. Renou, Y. Wang, S. Boreiri, S. Beigi, N. Gisin, and N. Brunner, Limits on correlations in networks for quantum and no-signaling resources, \textit{Phys. Rev. Lett.} \textbf{123}, 070403 (2019).

\bibitem{Luo2021}M.-X. Luo, Network configuration theory for all networks, arXiv:2107.05846, 2021.

\bibitem{Dual}F. Lad, G. Sanfilippo and G. Agr\'{o}, Extropy: Complementary Dual of Entropy, \textit{Statistical Science} \textbf{30}, 40-58 (2015).

\bibitem{YSL}V. S. Yepez, R. P. Sagar \& H. G. Laguna, Higher-order statistical correlations and mutual information among particles in a quantum well, \textit{Few-Body Systems} \textbf{58}, 158 (2017).

\end{thebibliography}
\end{document}